\documentclass[%
 reprint,
 amsmath,amssymb,
 aps,
prd,
]{revtex4-1}
\pdfoutput=1
\usepackage{amssymb,amsmath}
\usepackage{euscript}
\usepackage{mathtools} %include \prescript
\usepackage{enumitem}
\usepackage{graphicx}% Include figure files
\usepackage{dcolumn}% Align table columns on decimal point
\usepackage{bm}% bold math
\usepackage{hyperref}% add hypertext capabilities
\usepackage{array}
\newcommand{\Tz}{\overline{T}_{z}}
\newcommand{\Tzero}{\overline{T}_{0}}
\newcommand{\muK}{\mu\textnormal{K}}
\newcommand{\Dnu}{\EuScript{D}_{\nu_c}}
\newcommand{\Fnu}{\tilde{F}_\nu}
\newcommand{\Fnuc}{\tilde{F}_{\nu_c}}
\newcommand{\Bnu}{\tilde{B}_\nu}
\newcommand{\Bnuc}{\tilde{B}_{\nu_c}}
\newcommand{\Gij}[2]{\prescript{#1}{#2}{\mathcal{G}}}
 
\newcommand{\CEu}{\EuScript{C}}

\newcommand{\betahat}{\hat{\bm \beta}}
\newcommand{\gammahat}{\hat{\bm \gamma}}

\newcommand{\threeJ}[6]{
	\left(
	\begin{smallmatrix}
		#1 & #2 & #3 \\
		#4 & #5 & #6 
	\end{smallmatrix}\right)}

\begin{document}

\preprint{APS/123-QED}

\title{Kinetic Sunyaev Zeldovich effect in an anisotropic CMB model: \\ 
	measuring low multipoles of the CMB at higher redshifts using intensity and polarization spectral distortions}% Force line breaks with \\
%\thanks{}%

\author{Siavash Yasini}
\email{yasini@usc.edu} 
\author{Elena Pierpaoli}%
%\altaffiliation[Also at ]{}%Lines break automatically or can be forced with \\
 \email{pierpaol@usc.edu}
\affiliation{%
 Physics \& Astronomy Department, University of Southern California, Los Angeles, California 90089\\
}%

\date{\today}% It is always \today, today,

\begin{abstract}
 We present a novel mathematical formalism that allows to easily compute the expected kinetic Sunyaev Zeldovich (kSZ) signal in intensity and polarization due to an anisotropic primordial Cosmic Microwave Background (CMB). We derive the expected intensity and polarization distortions in the direction of non-moving galaxy clusters and then we generalize our calculations for non-zero peculiar velocity. We show that, in the direction of moving clusters, low CMB multipoles impose intensity and polarization spectral distortions with different frequency dependences. The polarization signal primarily probes the quadrupole moment of the CMB, with a significant contribution from the primordial dipole and octupole moments. For a typical cluster velocity of 1000 km/s, corrections to the quadrupole-induced polarization of a non-moving cluster are of the order of 2-10\% between 200-600 GHz, and depend on cluster's position on the sky, velocity magnitude and direction of motion.
 %  At 300 GHz, the sky area where the correction is greater than 10\% is about 20\%, although close to the galactic plane. 
 We also find that the angular dependence of the signal varies with frequency of observation. 
 The distinct frequency and angular dependences of the polarization induced by the primordial dipole and octupole can be exploited to 
 measure them despite other physical effects and foregrounds. Contrary to polarization, intensity distortions are affected by all the CMB multipoles, so they cannot be readily used to probe the low multipoles at higher redshifts. However, correlations between intensity and polarization signals, can be used to enhance the signal to noise ratio for the measurements of the primordial dipole, quadrupole and octupole. The more general calculation of the  aberration kernel presented in this work has applications reaching beyond the SZ cluster science addressed here.  For example, it can be exploited to the deboost/deaberrate CMB multipoles as observed in our local frame.
 
%\begin{description}
%\item[keywords]
%Cosmic Microwave Background, Quadrupole Induced Polarization, Sunyaev Zeldovich
%\end{description}
\end{abstract}

%\keywords{Sunyaev Zeldovich}%Use showkeys class option if keyword
\maketitle

%\tableofcontents

\section{\label{sec:level1}Introduction}

Intensity and polarization distortions of the Cosmic Microwave Background (CMB), induced in the direction of moving galaxy clusters, probe the peculiar velocity of galaxy clusters at higher redshifts. These distortions known as the intensity and polarization kinetic Sunyaev Zeldovich (kSZ) effects, are usually calculated assuming an isotropic incoming CMB towards moving clusters. Therefore, they do not completely reflect the contribution of the anisotropies to the observed signal. In an isotropic CMB model, the kSZ Intensity effect (kSZIn) \citep{Sunyaev:1980nv,Chluba2012c,Nozawa1998} probes the radial component of the cluster's peculiar velocity while the kSZ Polarization effect (kSZPol) \citep{Sunyaev:1972eq,Sunyaev:1980nv, Audit1999,Itoh:1998wv,Portsmouth2004a,Roebber:2013lra,Diego:2003dp} probes the transverse component. Since both effects are induced by the motion of cluster, they vanish in the direction of non-moving clusters. 

It has been shown that for non-moving galaxy clusters, the polarization-induced signal is proportional to the quadrupole moment ($\ell=2$) of the CMB observed at the cluster's location \citep{Hall2014,Sazonov1999,Baumann:2003xb,Bunn:2006mp,Cooray:2002cb,ramos2012cosmic,Liu:2016fqc,Challinor:1999yz,Seto:2005de}. Measuring this \textit{Temperature Induced Polarization} (TinPol) component can be used to infer the quadrupole moment of the CMB anisotropies at higher redshifts and reduce cosmic variance for this mode \citep{Kamionkowski1997a,Portsmouth2004c}. Independent measurements of the quadrupole moment can also potentially explain the low measured value of our local quadrupole compared to theory \citep{Smoot1992,PlanckXXIII2014,Tegmark2003}. 

By detailed calculation of the TinPol effect, we will show that in the direction of a moving cluster, the quadrupole is not the only mode that is reflected through the induced polarization distortion; all the other low multipoles of the CMB will have a contribution to the polarization signal with different frequency weights. In this paper we mainly focus on the contribution of the primordial octupole ($\ell=3$) and dipole ($\ell=1$) to the total induced polarization signal and  we investigate the possibility of their measurement at higher redshifts. Aside from the issue of cosmic variance, finding the octupole  moment at higher redshifts can help us determine if the apparent alignment between the quadrupole and octupole along the cosmic ``axis of evil" \citep{Tegmark2003,deOliveira-Costa2004,Copi2004,Liu:2016fqc} is just coincidental or if there is a fundamental reason for it that can be explained by physical laws. Also, the primordial dipole moment of the CMB at our location is overshadowed by the dipole generated by our motion in the CMB rest frame, therefore looking for it at the other locations in the universe would be a natural alternative to measure it.

The polarization distortion induced by the CMB anisotropies is a result of their \textit{Doppler and aberration leakage} into the quadrupole moment observed by the moving cluster. In order to calculate this leakage we use the aberration kernel formalism \citep{Challinor2002,Chluba2011,Dai2014,Kosowsky2010,Amendola2010,Yoho:2012am,Catena:2012hq,Menzies:2004vr,Aiola:2015rqa}, however, we generalize it by including the frequency dependence of the multipoles (which is typically either neglected or integrated over) and by allowing for a general peculiar velocity direction for the moving frame (which in previous works is always taken to be in the $\hat{\bm z}$ direction). These generalizations will reveal an interconnection between the direction of motion of the cluster and the frequency function of the observed signal which would have been concealed otherwise. We will also briefly discuss how the generalized aberration kernel can be employed in the local frame to deboost and deaberrate the observed CMB multipoles. 

Since the leakage of the low multipoles into the quadrupole is due to the motion of the cluster, naturally its induced polarization signal is proportional to the cluster's peculiar velocity $\beta=v/c$. The leakage of the quadrupole's first neighbors, the dipole and the octupole, is proportional to $\beta$, and that of its second neighbors, the monopole ($\ell=0$) and hexadecapole ($\ell=4$) to $\beta^2$, and so on. In the absence of temperature anisotropies our calculations naturally reduce to the kSZPol effect, which can be interpreted as the leakage of the CMB monopole into the quadrupole. Even though this effect is proportional to $\beta^2$, the monopole is larger than the other low multipoles by a factor of $\sim 10^5$ and consequently its induced polarization is the dominant effect for clusters with large peculiar velocities. Therefore, measurement of the polarization induced by the low multipoles requires identification and subtraction of the kSZ polarization in the direction of the cluster. Since the frequency dependence of the kSZ is different from the other low multipole-induced polarization signals, they can be easily separated in a multi-frequency survey.

 Apart from the kSZ polarization signal, the largest contribution to the TinPol after the quadrupole is due to the dipole and octupole. Since the leakage of these modes into the quadrupole is proportional to $\beta$, their polarization signals are expected to be relatively small. However we will show that the large frequency weights of the Doppler leakage of these modes amplify their induced polarization signals and even make them dominant over the quadrupole-induced polarization at high frequencies ($\nu\gtrsim400$ GHz). Furthermore, the amplitude of the quadrupole-induced polarization varies over the sky, independently of the peculiar velocities of the clusters, and vanishes in four different directions \citep{Sazonov1999,Hall2014,Amblard:2004yp}. Therefore in the areas around these four directions, where this signal is small, the primary source of polarization in the TinPol would be due to the dipole and octupole moments. 
 
 The low multipoles also induce a change in the intensity of the CMB observed in the direction of a galaxy cluster. This distortion, known as the blurring Sunyaev  Zedovich effect (bSZ) \citep{Hernandez-Monteagudo2010}, is present even in the direction of a non-moving cluster and is caused by scattering of the CMB photons out of the observer's line of sight. We will calculate this effect for an anisotropic CMB radiation, with corrections due to the motion of the cluster. For a moving cluster, the dipole moment of the bSZ effect will give us the well known kSZIn effect. We will refer to the quadrupole moment of bSZ effect as the \textit{Temperature-Induced Intensity} (TinIn) and show that its celestial distribution is highly correlated with the TinPol signal. There is also a strong correlation between the TinIn/TinPol effects at low redshifts and the local quadrupole moment of CMB observed at $z=0$, which can be exploited to enhance these signals. 
 
 The thermal Sunyaev Zeldovich effect (tSZ) \citep{Sunyaev1969} also induces an intensity and polarization distortion in the direction of the galaxy cluster. The tSZ intensity effect is typically much larger than its motion-induced counterpart, the kSZ intensity effect, and it has been measured for over a thousand galaxy clusters by the Planck satellite \citep{Aghanim:2015eva} as well as the ACT \citep{Hincks:2009vb} and SPT-SZ \citep{Bleem:2014iim} experiments. Both effects are proportional to the temperature monopole of the CMB and therefore are much larger than the low multipole-induced intensity effects. The tSZ polarization effect on the other hand, is sub-dominant to the kSZ polarization effect and hence comparable to the low multipole-induced polarization signals \citep{Sazonov1999}. Nevertheless, for the sake of simplicity we neglect the thermal effects throughout the calculations based on the assumption that they can be separated due to their distinct frequency functions. Since both tSZ intensity and polarization are proportional to the dimensionless temperature of the cluster $\theta_e=kT_e/m_ec^2$, corrections due to these effects can be treated perturbatively and simply added to the results of our calculations to first order in $\theta_e$. 
 
 kSZIn and kSZPol were first introduced in Refs. \citep{Sunyaev:1972eq} and \citep{Sunyaev:1980nv}. The TinPol effect was first suggested by Ref. \citep{Kamionkowski1997a} as a probe of the quadrupole mode at higher redshifts and has been examined in \citep{Sazonov1999,Portsmouth2004a,Hall2014,ramos2012cosmic,Baumann:2003xb,Bunn:2006mp,Challinor:1999yz}. Corrections due to thermal motion and relativistic effects in the absence of anisotropies have been studied in detail in Refs. \citep{Challinor:1999yz,Chluba2012c,Nozawa1998,Itoh:1998wv}. Our calculations include all of the above (except for the thermal effects) and extend the results to an anisotropic CMB model. 
 
 The outline of the paper is as follows: In \S\ref{sec:II} we derive the change in the intensity and polarization of the CMB in the direction of a non-moving cluster, induced by the primordial temperature anisotropies. In \S\ref{sec:III} we generalize the results for the case of a moving cluster, and show how the low multipoles of the CMB other than the quadrupole will contribute to the signal. We end with summary and conclusions in \S\ref{sec:IV}

\vspace{2em}
\section{\label{sec:II}Non-moving Galaxy Cluster}

\subsection{\label{sec:IIA}Geometry of the Problem and Notation}

\begin{figure}
	\centering
	\includegraphics[width=0.7\linewidth]{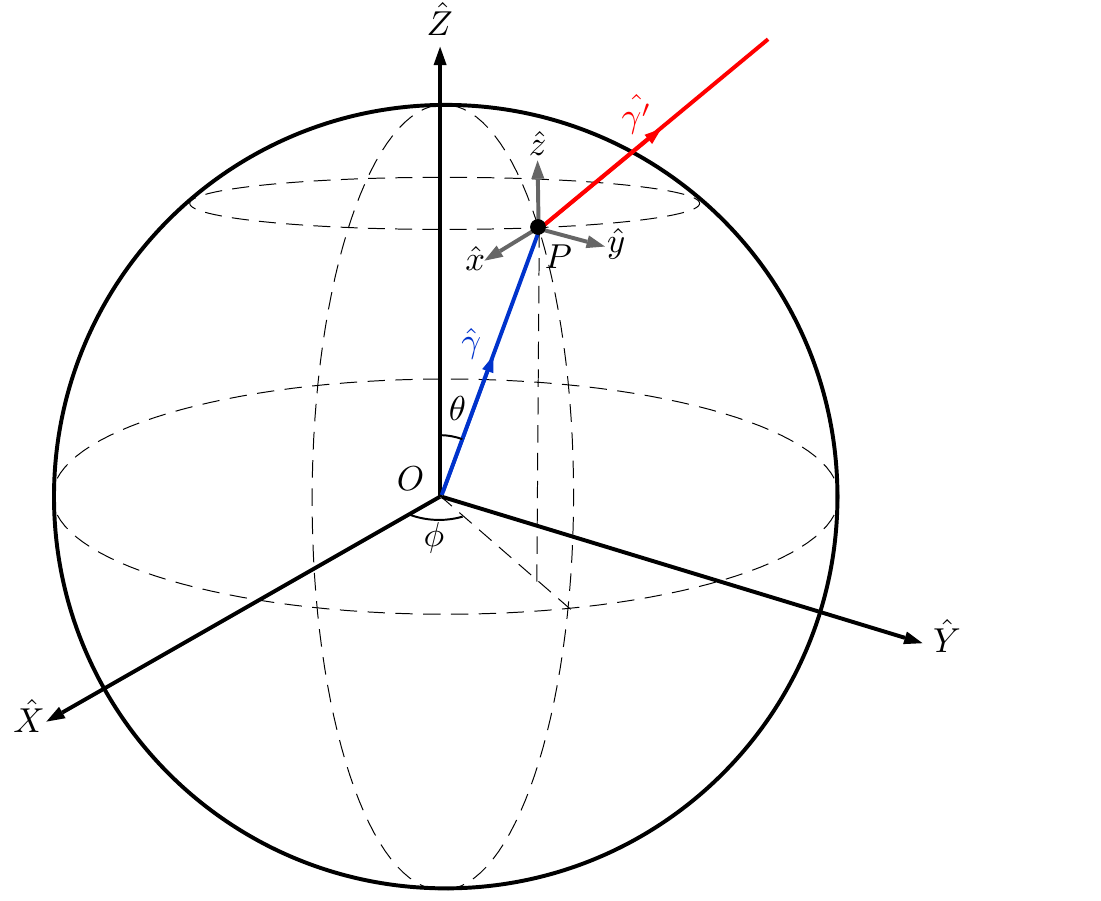}
	\caption{The CMB rest frame coordinate system $(\hat{\textbf{X}},\hat{\textbf{Y}},\hat{\textbf{Z}})$ is setup with the $\hat{\textbf{Z}}$ pointing towards the galactic north pole and $\hat{\textbf{X}}$ towards the galactic center. The cluster's frame  $(\hat{\bm x},\hat{\bm y},\hat{\bm z})$ is centered at the location of the cluster $P$ and initially aligned with the CMB rest frame. $\gammahat$ and $\gammahat'$ are respectively the line of sight vector of the observers at $O$ and $P$.}
	\label{fig:Figure1}
\end{figure}
For convenience we setup a coordinate system $(\hat{\textbf{X}},\hat{\textbf{Y}},\hat{\textbf{Z}})$ centered at the location of the observer $O$, with the $\hat{\textbf{Z}}$ and $\hat{\textbf{X}}$ respectively pointing towards the north galactic pole and the galactic center. We indicate the angular coordinates of the center of the cluster $P$ at distance $r$ from the origin $O$, with a line of sight vector $\gammahat=(\theta,\phi)$. The relationship between these angles and the galactic coordinates in radians is $(l,b)=(\phi,\pi/2-\theta)$. We set up the cluster's rest frame $(\hat{\bm x},\hat{\bm y},\hat{\bm z})$ at $P$ and initially align it with the observer's coordinate system (see figure \ref{fig:Figure1}). Throughout the calculations we keep the orientation of the observer's coordinate system $(\hat{\textbf{X}},\hat{\textbf{Y}},\hat{\textbf{Z}})$ fixed and apply the rotations and boosts only to the cluster's frame $(\hat{\bm x},\hat{\bm y},\hat{\bm z})$. The line of sight vector of the cluster will be denoted by $\gammahat'=(\theta',\phi')$. The actual propagation vectors of the incoming photons towards the cluster P and the outgoing ones towards the observer at $O$ are respectively $-\gammahat$ and $-\gammahat'$. In the paper $h$, $k$ and $c$ denote the Planck's constant, Boltzmann's constant and the speed of light. We also use $\sum\nolimits_{a,b,...,z}^{\alpha,\beta,...\zeta}$ as shorthand notation for $\sum\limits_{a=-\alpha}^{\alpha} \sum\limits_{b=-\beta}^{\beta}...\sum\limits_{z=-\zeta}^{\zeta}$. A list of all the abbreviations used in the paper is provided in table \ref{tab:1} (appendix \ref{sec:appC}).

\subsection{\label{sec:IIB}Intensity and Polarization Induced by a Non-moving Cluster}
\subsubsection{\label{sec:IIB1}The Radiative Transfer Equation}

 The specific intensity and polarization of the CMB are commonly described using the set of Stokes parameters $(I_\nu,Q_\nu,U_\nu)$ \citep{Chandrasekhar1950}. The change in these parameters due to Thomson scattering with the electrons in the cluster is conveniently expressed in terms of the radiative transfer equation \citep{Rybicki1986}

\begin{equation}\label{delta_I}
	\frac{\Delta I_\nu(\gammahat) }{\Delta\tau}  = \frac{3}{16\pi} \int I_\nu(\gammahat') (1+\cos^2\theta_{sc})\text{d}^2\gammahat' -I_\nu(\gammahat), 
\end{equation} 

\begin{equation}\label{delta_Q+iU}
	\frac{\Delta (Q_\nu\pm iU_\nu)(\gammahat) }{\Delta\tau} = \frac{3}{16\pi} \int 
	I_\nu(\gammahat') \sin^2\theta_{sc}~e^{\pm 2i\phi_{sc}}
	\text{d}^2 \gammahat',
\end{equation}
 where $\theta_{sc}$ and $\phi_{sc}$ are the polar and azimuthal angles of scattering between $\gammahat$ and $\gammahat'$, and $\tau=\int n_e \sigma_T ds$ is the optical depth of the cluster with respect to Thomson scattering. Here $n_e$ is the number density of the electrons, $\sigma_T$ is the Thomson cross section and $s$ is the length of the cluster along the line of sight. 
 
The thermal motion of the electrons (which causes the tSZ distortion) and multiple scattering effects \citep{Chluba2013a, Chluba2013b} are neglected to simplify the calculations. The scattering events are assumed to be elastic with no energy transfer between the electrons and photons. In the limit of cold clusters, since the scatterings do not change the bulk motion of the cluster, it is safe to use Thomson scattering \citep{Portsmouth2004a}. For high temperature clusters, since the electrons upscatter the CMB photons, the elasticity assumption does not hold anymore and Compton scattering should be used instead \citep{Portsmouth2004b}. The tSZ corrections in this limit are of the order of the perturbative parameter $\theta_e\equiv kT_e/m_ec^2\approx0.01$ for a typical cluster, so if calculated, they can be linearly added to equations \eqref{delta_I} and \eqref{delta_Q+iU}. The initial polarization of the CMB is also neglected, but the polarization-induced effects also linearly couple to equations \eqref{delta_I} and \eqref{delta_Q+iU}, so the extra terms can be simply added at any point. It is important to mention that although the polarization-induced effects are typically smaller than the temperature-induced ones, due to their different spatial morphology they can be comparable or even dominant over them in certain directions over the sky. 

In order to integrate the radiative transfer equations for an anisotropic incoming intensity with the harmonics expansion 
 \begin{equation}\label{intensity_expansion}
	I_\nu(\gammahat') = \sum_{\ell=0}^{\infty}\sum\nolimits_{m}^{\ell} a_{\ell m}^{I}(\nu)~ Y_{\ell m}(\gammahat'),
 \end{equation}
 we rewrite the integrand of equations \eqref{delta_I} and \eqref{delta_Q+iU} in terms of spin-weighted spherical harmonics using the generalized addition theorem \citep{Hu:1997hp,Ng1997,Bars}
\begin{multline}\label{addition}
	_{s}Y_{\ell s}(\theta_{sc},0)e^{-is\phi_{sc}}=\\
	(-1)^s
	\sqrt{\frac{4\pi}{2\ell+1}}\sum\nolimits_{m}^{\ell}Y^*_{\ell m}(\gammahat')_{s}Y_{\ell m}(\gammahat).
\end{multline}
 After integrating over all the incoming photons $\gammahat'$ these equations simplify to 
 
 \begin{equation}\label{Delta_I_gamma}
 \frac{\Delta I_\nu(\gammahat) }{\Delta\tau}  =-\overbrace{ \delta I_{\nu}^{(1)}(\gammahat) }^{\text{bSZ}^{(1)}}-\overbrace{ \delta I_{\nu}^{(2)}(\gammahat) }^{\text{TinIn}}-\sum_{\ell=3}^{\infty}\overbrace{ \delta I_{\nu}^{(\ell)}(\gammahat) }^{\text{bSZ}^{(\ell)}},
 \end{equation}
 
 \begin{equation}\label{Delta_Q+iU_gamma}
 \frac{\Delta (Q_\nu\pm iU_\nu)(\gammahat) }{\Delta\tau} =-\overbrace{\delta(Q_\nu\pm iU_\nu)^{(2)}(\gammahat) }^{\text{TinPol}},
 \end{equation}
 with each term defined as 
 \begin{equation}\label{TinIn}
 \delta I_\nu^{(2)}(\gammahat) = \frac{9}{10}\sum\nolimits_{m}^{2} a^{I}_{2 m}(\nu) ~Y_{2 m}(\gammahat),
 \end{equation}

 \begin{equation}\label{TinPol}
 \delta (Q_\nu\pm iU_\nu)^{(2)}(\gammahat)= \frac{\sqrt{6}}{10}\sum\nolimits_{m}^{2}  a^{I}_{2m}(\nu) ~_{\mp 2}Y_{2m}(\gammahat),
 \end{equation}
 and for $\ell \neq 2$ 
 \begin{equation}\label{bSZ}
 \delta I_\nu^{(\ell)}(\gammahat) = \sum\nolimits_{m}^{\ell} a^{I}_{\ell m}(\nu) ~Y_{\ell m}(\gammahat).
 \end{equation}

The right hand side (RHS) of equation \eqref{TinPol} is induced by the photons that scatter into the line of sight of the observer, while the RHS of equation \eqref{TinIn} is a combination of the photons that  scatter into and out of the line of sight. We refer to these terms as the \textit{Temperature-Induced Polarization} (TinPol) and \textit{Temperature-Induced Intensity} (TinIn) effects. Notice that both these effects depend on the quadrupole of the CMB as observed by the cluster and therefore are highly correlated with each other. Equation \eqref{bSZ} is basically the harmonic expansion of the photons that scatter out of the line of sight. We will refer to these terms as the blurring Sunyaev Zeldovich effect (bSZ) \citep{Hernandez-Monteagudo2010}. Although the temperature quadrupole-induced intensity $\delta I_\nu^{(2)}$ is a combination of both in-scattered and out-scattered photons, since the frequency dependence of these scattering events are identical they are not distinguishable from each other. Therefore the overall effect is written as one single term, which we have called the TinIn effect. However, it is important to remember that the nature of this effect is different from the bSZ, which is only caused by out-scattering of the photons. 

 Note that the monopole term that survives the radiative transfer integral, has been canceled out with the monopole term of the bSZ expansion. This means that in a completely isotropic CMB model where all the $a_{lm}$s for $l>0$ vanish, $\Delta I_\nu(\gammahat) $ and $\Delta (Q_\nu\pm iU_\nu)(\gammahat)$ will be equal to zero. This is because in an isotropic model the number of scatterings that deflect the photons towards and out of the line of sight are equal to each other. However it is evident from equations \eqref{TinIn} and \eqref{TinPol} that after including the anisotropies, the quadrupole moment of the incident radiation induces a change in the observed intensity and polarization of the CMB in the direction of the cluster $\gammahat$. It is important to note that there is an implicit location/redshift dependence in all the $a_{\ell m}$s on the RHS of these equations. Therefore the TinIn and TinPol signals give us access to the observed quadrupole moment of the CMB at the redshift of the cluster $z$ as first noticed in \citep{Kamionkowski1997a}.

\subsubsection{\label{sec:IIB2}Separating Frequency Dependence from Spatial Morphology}
 
 The specific intensity of the CMB has an almost perfect black body spectrum which can be described as $I_\nu(\gammahat')=B_\nu (T(\gammahat'))$ where $B_\nu$ is the Planck function defined as $B_\nu(T)\equiv\frac{2h\nu^3}{c^2}\frac{1}{e^{h\nu/kT}-1}$ and $T(\gammahat')$ is the thermodynamic temperature of the CMB in the $\gammahat'$ direction. If we assume that the frequency spectrum of the CMB  is isotropic (by negliecting anisotropic effects such as patchy reionization \citep{Santos2003,Dore2007}), we only need a single number, namely the thermodynamic temperature of the blackbody $T$, and an appropriate frequency function (here the Planck Function $B_\nu$) to describe the intensity of the CMB in every direction $\gammahat$. Consequently the intensity multipole coefficients $a^{I}_{\ell m}(\nu)$, can be expressed as the combination of a frequency function which describes the spectrum of the anisotropies, and the thermodynamic temperature multipole coefficients $a^{T}_{\ell m}$ defined as 

\begin{equation}\label{temperature_expansion}
T(\gammahat') = \sum_{\ell=0}^{\infty}\sum\nolimits_{m}^{\ell}~a^{T}_{\ell m}~ Y_{\ell m}(\gammahat').
\end{equation} 
Using the Planck frequency function it is easy to show that the relationship between these coefficients up to first order in temperature anisotropies is (see appendix \ref{sec:appA} and \citep{Chluba:2012gq})

\begin{subequations}\label{alm(I)_to_alm(T)}
	\begin{align}
	a^{I}_{00}(\nu)&=\tilde{B}_\nu(\overline{T})~a^{T}_{00},\\
	a^{I}_{\ell m}(\nu)&=\tilde{F}_\nu(\overline{T})~a^{T}_{\ell m}\quad(l>0),
	\end{align}
\end{subequations}
 where $\overline{T}=\frac{a^{T}_{00}}{2\sqrt{\pi}}$ is the thermodynamic temperature of the CMB monopole, $\tilde{B}_\nu(\overline{T})\equiv \overline{T}^{-1} B_\nu(\overline{T})$ and $\tilde{F}_\nu(\overline{T})\equiv \overline{T}^{-1}B_\nu(\overline{T})f(h\nu/k\overline{T})$ are  respectively the frequency functions of the monopole and higher multipoles of the anisotropies, with $f(x)\equiv\frac{xe^{x}}{e^{x}-1}$. The tilde over the frequency functions denotes the normalization by the temperature monopole $\overline{T}$. The $a^{I}_{\ell m}(\nu)$ coefficients have units of radiative intensity ($\text{W/m}^2\text{Hz sr}=10^{26}$Jy/sr) and $a^{T}_{\ell m}$s have units of temperature (Kelvin). 
 
 Using equations \eqref{alm(I)_to_alm(T)} in \eqref{TinIn} and \eqref{TinPol} will give us the advantage of separating the frequency dependence of the TinIn and TinPol signals from their spatial morphology over the sky, simply described by the $a^{T}_{\ell m}$s. For a non-moving cluster this separation is trivial, but in the case of a moving cluster it will help us to easily calculate the frequency weights of the leakage of the low multipoles into the quadrupole due to the Doppler effect. After substitution, equations \eqref{TinIn} and \eqref{TinPol} can be rewritten as
 
 \begin{equation}\label{TinIn2}
 \delta I_\nu^{(2)}(\gammahat) = \frac{9}{10}\Fnu(\Tz)\sum\nolimits_{m}^{2} a^{T_z}_{2 m} ~Y_{2 m}(\gammahat),
 \end{equation}
 
 \begin{equation}\label{TinPol2}
 \delta (Q_\nu\pm iU_\nu)^{(2)}(\gammahat)= \frac{\sqrt{6}}{10}\Fnu(\Tz)\sum\nolimits_{m}^{2}  a^{T_z}_{2m} ~_{\mp 2}Y_{2m}(\gammahat),
 \end{equation}
 and similarly for the bSZ terms (eq. \eqref{bSZ})
 \begin{equation}\label{bSZ2}
 \delta I_\nu^{(\ell)}(\gammahat) = \Fnu(\Tz)\sum\nolimits_{m}^{\ell} a^{T_z}_{\ell m} ~Y_{\ell m}(\gammahat),
 \end{equation}
where the  $a^{T_z}_{\ell m}$s are the harmonic coefficients of the temperature anisotropies of the CMB observed by the cluster at redshift $z$ and $\Tz$ is the average temperature at that redshift. As we will see in \S \ref{sec:III}, the bSZ term $\delta I_{\nu}^{(1)}$ will give rise to the kSZ intensity distortion \citep{Sunyaev:1980nv}. 
 
 \begin{figure}[t]
 	\centering
 	\text{217 GHz}\par \medskip
 	\
 	\includegraphics[width=0.9\linewidth]{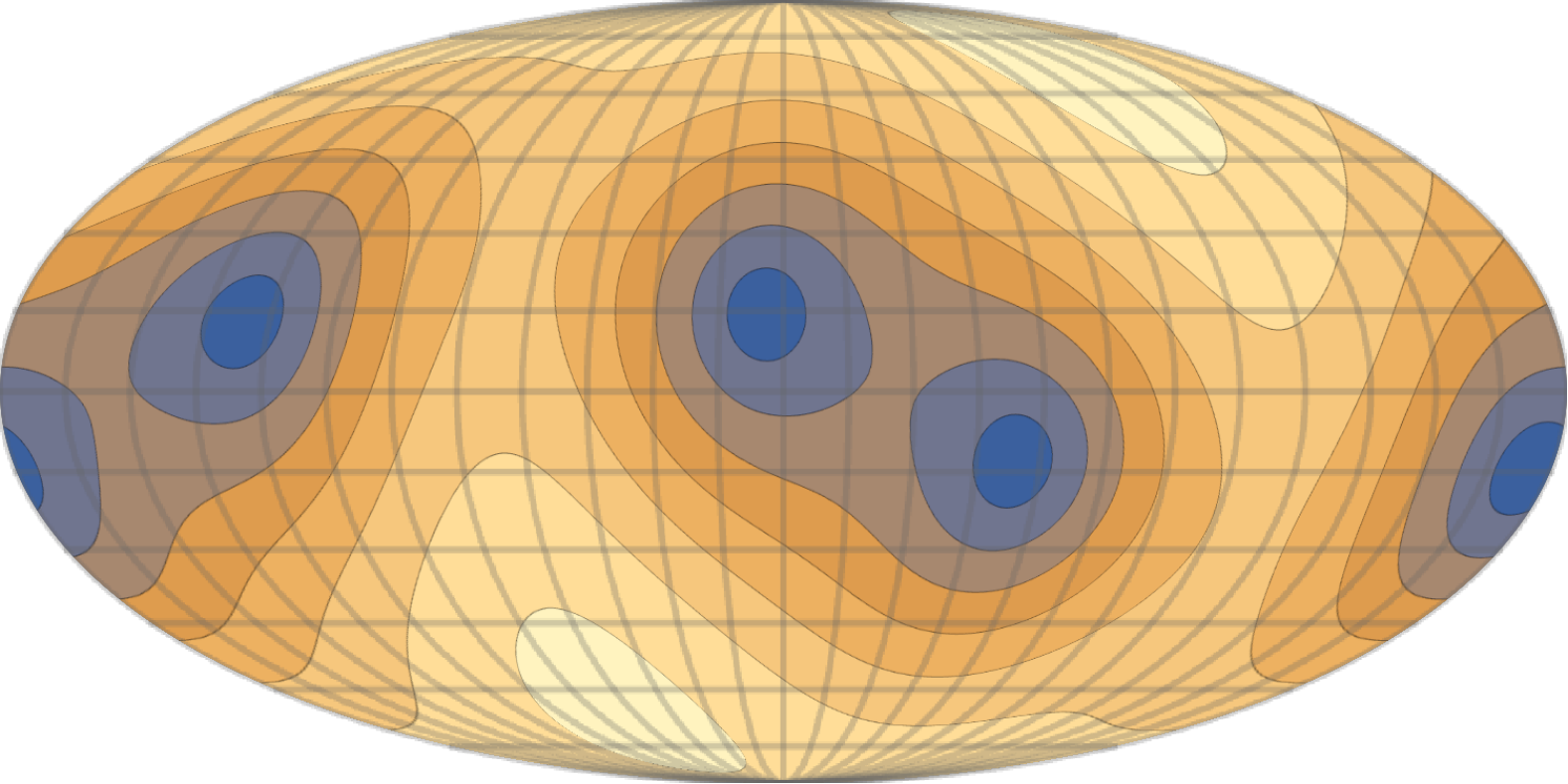}
 	\includegraphics[width=0.9\linewidth]{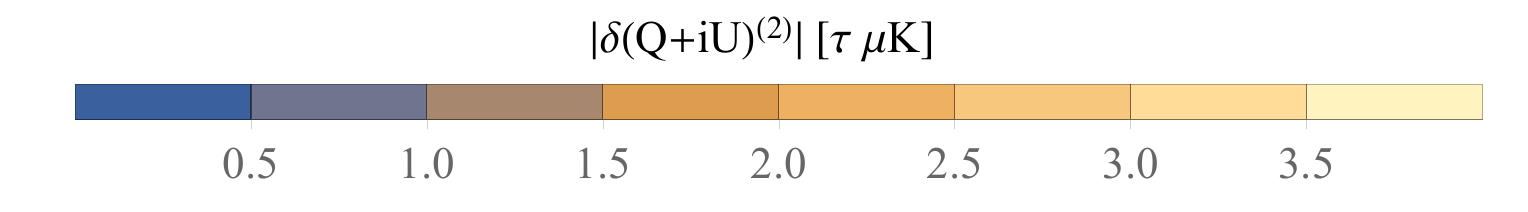}
 	\text{217 GHz}\par \medskip
 	\includegraphics[width=0.9\linewidth]{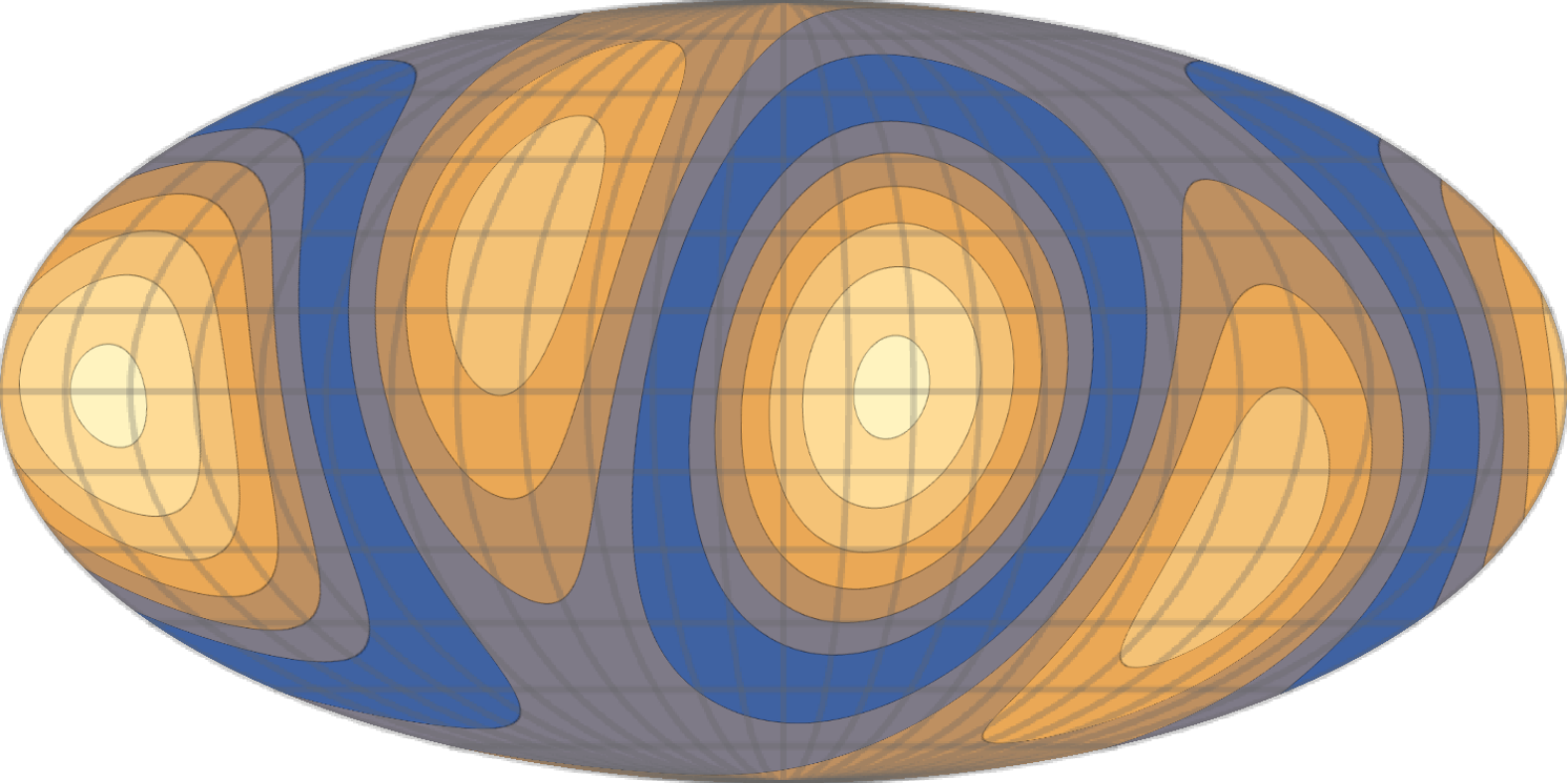}
 	\includegraphics[width=0.9\linewidth]{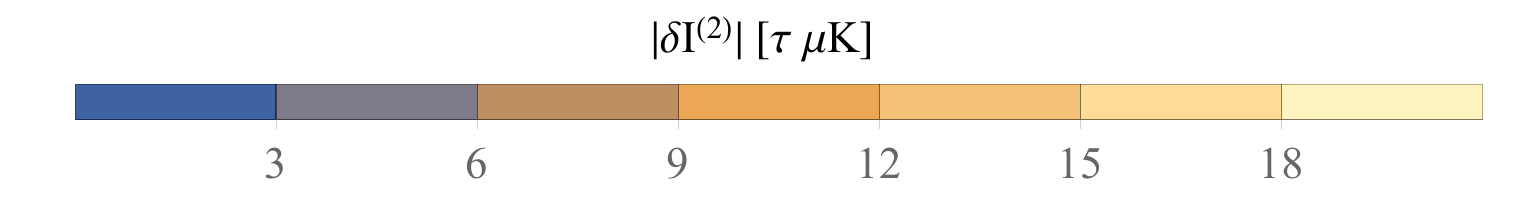}
 	\caption{Mollweide projection contour maps of the TinPol (\textit{top}) and TinIn (\textit{bottom}) at 217 GHz at $z=0$ in galactic coordinates.  The quadrupole moment of the CMB observed by the clusters is estimated using Planck's SMICA map. The TinPol signal vanishes at 4 different directions and reaches the maximum value of $1.7\tau \text{kJy/sr}~(3.6 \muK)$at $(l,b)=(-113.1^{\circ},-63.3^{\circ})$ and $(67.615^{\circ},67.873^{\circ})$. The TinIn signal vanishes over two large rings and has the maxuimum value of $9.1\tau\text{kJy/sr}~(18.7 \muK)$ at $(l,b)=(-23.6^{\circ},0.4^{\circ})$ and $(100.4^{\circ},-0.5^{\circ})$. }
 	\label{fig:TIT}
 \end{figure}

Here, since the frequency dependence of these signals are the same as that of the CMB temperature anisotropies, we can also write the distortions in terms of the thermodynamic temperature defined as

\begin{equation}\label{delta_temperature}
\delta T(\gammahat)\equiv \frac{\delta I_{\nu}(\gammahat)}{\Fnu(\Tz)}, 
\end{equation}
as is common practice in tSZ and kSZ calculations. A similar conversion can be used for the polarization parameters $Q_T$ and $U_T$ as well. After plugging in equations \eqref{TinIn2}-\eqref{bSZ2} into \eqref{delta_temperature}, it would be evident that the thermodynamic temperature (and polarization) distortions induced by the quadrupole in the direction of non-moving clusters are not frequency-dependent. However, this is not the case for the distortions induced by the motion of the cluster. As will be shown later, due to the addition of the motion-induced effects, the total change in intensity and polarization will have a nontrivial frequency function that cannot be described by the blackbody spectrum anymore. Therefore, although equation \eqref{delta_temperature} can be used to convert all the results in terms of temperature fluctuations, in order to see the complete frequency dependence of the signal and for the sake of conceptual clarity we refrain from using this conversion throughout the calculations. 

In order to get an estimate of the amplitude of the TinPol and TinIn signals for non-moving clusters, we set the bSZ terms equal to zero and use the numerical values of the  $a^T_{2m}$s from the de-boosted Planck SMICA temerature map \citep{Polastri:2015rda} as an estimate of the quadrupole moment that a galaxy clusters at $z\approx0$ would observe. Figure \ref{fig:TIT} shows the angular distribution of these signals over the whole sky. Both signals have similar frequency dependencies but they have different angular distributions over the sky. The TinPol and TinIn signals reach the maximum values of $\Delta (Q+iU)^{\text{max}}_{\nu} = 3.6 \Fnu(\Tzero) \tau \muK$ and $\Delta I^{\text{max}}_{\nu} =18.6 \Fnu(\Tzero)\tau \muK$. At $\nu=\text{217 GHz}$ where the tSZ signal vanishes, these values correspond to $\Delta (Q+iU)^{\text{max}}_{\text{217 GHz}} \approx 1.7 \tau$kJy/sr and  $\Delta I^{\text{max}}_{\text{217 GHz}} \approx 9.1 \tau$kJy/sr. The reader should keep in mind that at higher redshifts the amplitude and angular dependence of the signal will be different due to the change in $a^T_{2m}$ coefficients. However, this change is very small up to $z\sim1$ \cite{Seto:2005de} so for close by clusters we expect the distortion maps to be highly correlated with the ones in figure \ref{fig:TIT}.
 
 In actual observations, the intensity distortion map in the direction of galaxy clusters contains both TinIn and bSZ effects. Since these effects have the same frequency dependence, it is not easy to separate them without prior knowledge of the temperature anisotropies at higher redshifts. Therefore, the TinIn effect would not be directly observable as opposed to the TinPol. However, by looking at the definition of the TinIn effect in equation \eqref{TinIn2} and its angular distribution in figure \ref{fig:TIT}, it is evident that at $z=0$ this effect is equal to our local quadrupole rescaled by a factor of $0.9\tau$ (compare with figure 17 of \citep{PlanckXXIII2014}). Therefore, we expect the cluster intensity distortion map over the whole sky at low redshifts to be highly correlated with the local quadrupole observed at $z=0$. This correlation can be used to amplify the TinIn signal and separate it from bSZ for clusters at low redshifts.  Moreover, using the definition of the spin-2 spherical harmonics through \citep{goldberg1967spin,Hu:1997hp}
 \begin{equation}
 _{\pm2}Y_{\ell m}(\gammahat)\equiv \sqrt{\frac{(\ell\mp2)!}{(\ell\pm 2)!}}(-1)^{\pm2}\eth^{\pm2} Y_{\ell m}(\gammahat)
 \end{equation}
 with the operators $\eth^{\pm2}$ defined as 
 
  \begin{multline}
  \eth^{\pm2} \equiv \frac{\partial^2}{\partial\theta^2}-\cot(\theta)\frac{\partial}{\partial\theta} \pm\frac{2 i}{\sin(\theta)}\frac{\partial^2}{\partial\theta \partial\phi} \\ -\cot(\theta)\frac{\partial}{\partial\phi}-\frac{1}{\sin^2(\theta)}\frac{\partial^2}{\partial\phi^2},
  \end{multline}
  it is easy to see that the $\eth^{\pm2}$ derivative of the TinIn map in equation \eqref{TinIn2} is just a rescaling of the TinPol in \eqref{TinPol2}. Therefore the second derivative of the quadrupole moment of the intensity distortion map is strongly correlated with the polarization distortion map. Furthermore, since TinIn itself is correlated with the local quadrupole at low redshifts, the second derivative of the quadrupole map is also highly correlated with the TinPol. In other words the second derivative of the TinIn map at low redshifts (bottom panel of figure \ref{fig:TIT}) and the local quadrupole, both reproduce a rescaled version of the TinPol map (top panel of figure \ref{fig:TIT}) and can be used to enhance the signal to noise ratio for this effect. 
  
\section{\label{sec:III}Moving Galaxy Cluster}

 We showed that the intensity and polarization-induced in the direction of a galaxy cluster are proportional to the quadrupole moment that the cluster observes in its location, so by measuring these induced signals we can infer the $a^{T_z}_{2 m}$s at that redshift. However if the cluster is moving, since equations \eqref{TinIn} and \eqref{TinPol} are not Lorentz invariant, the observed quadrupole moment in the cluster's frame ---which sources the TinIn and TinPol signals--- will not be equal to the primordial quadrupole moment in the CMB frame anymore. Hence, in order to correctly use these equations in the cluster's frame, we need to find the proper transformation that links the former to the latter.
 
 Using the generalized aberration kernel formalism, we will show that the quadrupole moment observed in the moving cluster's frame is not only proportional to the primordial quadrupole in the CMB rest frame, but also to all the other low multipoles of the temperature anisotropies. We will see that the multipole moments observed in a moving frame have contributions from their nearby multipoles in the CMB rest frame as a result of the aberration and Doppler effects. The aberration effect is only a geometrical effect and does not change the frequency of the photons, therefore the \textit{aberration leakage} of the multipoles into each other is frequency-independent. The Doppler effect, on the other hand, depends both the frequency and angle of the incoming photons, which makes the \textit{Doppler leakage} of the multipoles into each other frequency-dependent. In a moving frame, the frequency spectrum of the CMB will still be that of a blackbody in every single direction, however, the spectrum of the multipoles will be distorted due to mixing of these blackbodies \citep{Chluba:2012gq,Chluba:2004cn}. Therefore, the combination of the Doppler and aberration effects changes the frequency spectrum of the observed multipole coefficients in the frame of the moving cluster. This frequency dependence is always concealed in calculations of the kernel of the transformation between the two frames, because they are typically carried out in terms of the temperature coefficients $a^{T}_{\ell m}$ or the frequency integrated coefficients $a^{I}_{\ell m}=\int a^{I}_{\ell m}(\nu)\text{d}\nu$ \citep{Challinor2002,Chluba2011,Kosowsky2010,Amendola2010}. Using the frequency-dependent intensity coefficients, however, we will find the frequency functions with which different multipoles leak into each other. These frequency functions will allow us to distinguish different multipole-induced effects from one another in a multi-frequency survey.

 In this section, first we calculate the general expression for the observed intensity multipole coefficients $a^{I_{c}}_{\ell m}$ in the frame of a cluster moving in an arbitrary direction, as a function of the primordial $a^{I}_{\ell m}$s in the CMB rest frame. Then we will show that the quadrupole observed by the cluster, which is the only mode that is reflected through the TinIn and TinPol signals, will have contributions from the primordial dipole, quadrupole, octupole, hexadecapole etc. with different frequency functions. Therefore all these low multipoles are observable through the intensity and polarization signals induced in the direction of the cluster (equations \eqref{TinIn} and \eqref{TinPol}). Since the frequency weights of the Doppler leakage of the low multipoles into the quadrupole are related to the higher order derivatives of the blackbody frequency spectrum, they will amplify the overall TinIn and TinPol signals (see appendix \ref{sec:appA}). 
 
 There is also a leakage from the temperature monopole into the quadrupole which leads to the kSZ intensity and polarization effects. Due to the large amplitude of the monopole relative to the anisotropies, its induced intensity and polarization signals are typically larger than the ones induced by the other low multipoles. However, since the frequency dependence of these effects are not the same, we may be able to distinguish them from each other. We avoid any frequency integrations over the $a^I_{\ell m}$s throughout the calculations to show this feature clearly. We will also do a comparison between kSZ and the low multipole-induced intensity and polarization effects. In the following section the subscript $c$ corresponds to variables in the cluster's frame. 

 \subsection{\label{sec:IIIA}CMB Intensity Observed by a Moving Cluster}

 In order to calculate the multipole coefficients $a^{I_c}_{\ell m }(\nu)$ of the CMB observed by the cluster in terms of the $a^{I}_{\ell m}(\nu)$ coefficients in the CMB rest frame we take advantage of the Lorentz invariance of the quantity $I_\nu/\nu^3$ to relate the intensities in the different frames. For a cluster that is moving with the peculiar velocity vector $\vec{\bm v}_{pec}=\vec{\bm \beta}c$ in the CMB rest frame $(\hat{\textbf{X}},\hat{\textbf{Y}},\hat{\textbf{Z}})$, the observed intensity in $\gammahat_c=(\theta_c,\phi_c)$ direction at frequency $\nu_c$ can be written as \citep{Challinor2002,Rybicki1986}
\begin{equation}\label{I_prime}
	I_{\nu_c}( \gammahat_{c})=\Big(\frac{\nu_{c}}{\nu_{cmb}}\Big)^3 I_{\nu_{cmb}}( \gammahat_{cmb}).
\end{equation}
 Here $\nu_{cmb}$ and $\gammahat_{cmb}$ are the frequency and the line of sight vector of the incoming photon in the CMB rest frame. The observed frequency and direction vector observed in the Lorentz boosted cluster frame $(\hat{\bm{x}},\hat{\bm{y}},\hat{\bm{z}})$ are given by
 \begin{equation}\label{doppler_freq}
 	\nu_c=\Big(\frac{1+\beta \mu}{\sqrt{1-\beta^2}}\Big)\nu_{cmb},
 \end{equation}
 	
 \begin{equation}\label{aberration}
 	\gammahat_c=\Big(\frac{(1-\sqrt{1-\beta^2})\mu+\beta}{1+\beta \mu}\Big )\betahat+\Big (\frac{\sqrt{1-\beta^2}}{1+\beta \mu}\Big)\gammahat_{cmb}.
 \end{equation}
 Here $\mu=\gammahat_{cmb}.\betahat$, $\beta=|\vec{\bm \beta}|$ and $\betahat=\vec{\bm \beta}/\beta$ which can be denoted as $(\theta_\beta,\phi_\beta)$ in the CMB frame. Using the spherical harmonic expansion of equation \eqref{intensity_expansion} on both sides of equation \eqref{I_prime}, we write the multipole coefficients observed in the cluster's frame $a^{I_c}_{\ell' m'}(\nu_c)$, in terms of the coefficients in the CMB rest frame $a^{I}_{\ell m}(\nu_{cmb})$
 
 \begin{multline}\label{alm_cluster}
 a^{I_c}_{\ell' m'}(\nu_c) =\\
 \sum_{\ell=0}^{\infty}\sum\nolimits_{m}^{\ell} \int \Big(\frac{\nu_c}{\nu_{cmb}}\Big)^3 a^{I}_{\ell m}(\nu_{cmb}) Y_{\ell m}(\gammahat_{cmb}) Y^*_{\ell' m'}(\gammahat_c) d^2 \gammahat_c.
 \end{multline}
 This expression relates the harmonic coefficients of the observed multipoles in the two frames, so we can use it to find the quadrupole in the cluster's frame $a^{I_c}_{2 m'}$ in terms of all the primordial $a^I_{\ell m}$s in the CMB rest frame. In the next two subsections we use the inverse of equations \eqref{doppler_freq} and \eqref{aberration} to expand $a^{I}_{\ell m}(\nu_{cmb})$ and $ Y_{\ell m}(\gammahat_{cmb}) $ in terms of their cluster frame counterparts so that we can integrate the RHS of this expression.
 
 \subsubsection{\label{secIIIA1}Doppler effect}
 
 For typical values of $\beta$ the Doppler shift  in equation \eqref{doppler_freq} is small. This allows us to use the following Taylor expansion:
 
 \begin{multline}\label{Doppler_expand}
 \left(\frac{\nu_c}{\nu_{cmb}}\right)^3a^{I}_{\ell m}(\nu_{cmb})=
 \Dnu^{(00)}a^{I}_{\ell m}(\nu)\\
 +\beta~\Dnu^{(11)}a^{I}_{\ell m}(\nu)\sum\nolimits_{n}^{1}\frac{4\pi}{3}Y_{1n}(\gammahat_c)Y^*_{1n}(\betahat)\\
 +\beta^2\Dnu^{(22)}a^{I}_{\ell m}(\nu)\sum\nolimits_{n}^{2}\frac{4\pi}{5}Y_{2n}(\gammahat_c)Y^*_{2n}(\betahat)\\
 +\beta^2\Dnu^{(20)}a^{I}_{\ell m}(\nu)+O(\beta^3).
 \end{multline} 
  where the differential operators $\Dnu^{(kj)}$ are defined in the appendix (see \ref{sec:appB1a}). The superscripts are chosen such that individual terms on the RHS are in $\beta^k\Dnu^{(kj)}Y_{j n}(\gammahat_c)$ format; they are simply labels to distinguish these operators and they have no tensorial meaning. Each operator $\Dnu^{(kj)}$ consists of frequency derivatives up to the order $\partial_{\nu_c}^k\equiv \partial^k / \partial\nu^k |_{ \nu = \nu_c}$. The trivial operator $\Dnu^{(00)}$ only changes the argument of its following function from $\nu$ to $\nu_c$ and is defined for consistency in notation. In this equation, since the Doppler factor $(\nu_c/\nu_{cmb})^3$ only depends on powers of $\mu_c=\gammahat_c.\betahat$, we have written the expansion in terms of Legendre polynomials and separated the $\gammahat_c$ and $\betahat$ dependence using the addition theorem
 
 \begin{equation}\label{addition_theorem}
 P_N(\gammahat_c.\betahat)=\frac{4\pi}{2N+1}\sum\nolimits_{n}^{N} Y_{N n}(\gammahat_c) Y^*_{N n}(\betahat).
 \end{equation}
 Equation \eqref{Doppler_expand} introduces the leakage of the intensity harmonic coefficients $a^I_{\ell m}$s due to the Doppler effect. The second line of this equation shows the Doppler leakage of the first neighbors of the $\ell^{\text{th}}$ multipole into it. Using the addition property of the spherical harmonics $ Y_{1m}\times  Y_{\ell m}  \propto Y_{(\ell \pm 1)m}$, it is easy to see that in the absence of the aberration effect, after substituting this equation back into \eqref{alm_cluster} the observed $a^{I_c}_{\ell' m'}$s in the moving frame have contributions from $a^I_{(\ell'\pm1) m'}$ coefficients in the CMB frame up to first order in $\beta$. Similarly, the third line in equation \eqref{Doppler_expand} brings in contributions from $a^I_{(\ell'\pm2) m'}$ to second order in $\beta$. 
 
 The angle dependence of this expansion is due to the fact that the ratio of the observed frequencies in each frame depends on the observation angle.  To first order in $\beta$ the Doppler factor $(\nu_c/\nu_{cmb})^3$ only depends on the cosine of the angle between the incoming photons and the direction of motion of the cluster so it will distort the $\ell^{\text{th}}$ multipole in the moving frame by a factor of $\mu_c=P_1(\gammahat_c.\betahat)$. Therefore it is maximal when the line of sight is parallel to the direction of motion and is zero when it is perpendicular, as there is no Doppler shift to first order in $\beta$ in these directions. It is evident from equation \eqref{Doppler_expand} that the Doppler effect  distorts both the frequency spectrum of the observed multipoles (\textit{frequency leakage}), and their angular spectrum (\textit{geometrical leakage}). Keep in mind that the intensity in every direction is still a blackbody and the Doppler effect only distorts the spectrum of the multipoles. In particular it draws in the first neighbors of  each multipole $(\ell\pm1)$ with the frequency operator $\Dnu^{(11)} (\propto \partial_{\nu_c}^1 )$ to first order in $\beta$, the second neighbors $(\ell\pm2)$ with the frequency operator $\Dnu^{(22)} (\propto \partial_{\nu_c}^2 )$ to second order in $\beta$, and so on.

\subsubsection{\label{secIIIA2}Aberration effect}

 We use equation \eqref{aberration} to expand $Y_{\ell m}(\gammahat_{cmb})$ in terms of $Y_{\ell m}(\gammahat_{c})$ (see appendix \ref{sec:appB1b})
 
 \begin{figure}\label{leakage}
 	\centering
 	\includegraphics[width=1.05\linewidth ]{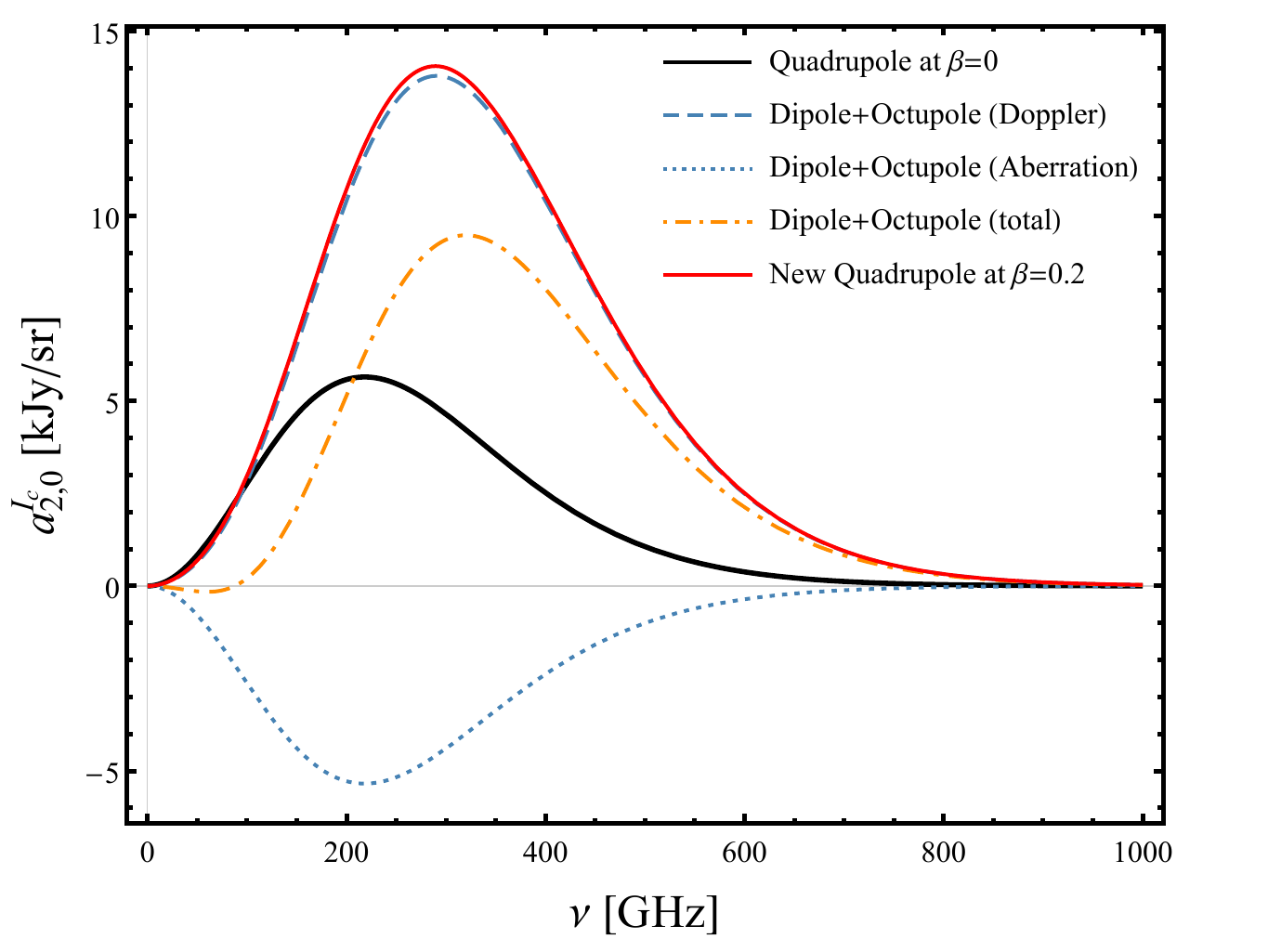}
 	\caption{Leakage of the dipole and octupole moments of the primordial CMB into the $a^{I_c}_{2,0}(\nu)$ coefficient of the quadrupole observed by a cluster that is moving in the $-\hat{\bm z}$. The frequency functions of the Doppler (\textit{dashed}), aberration (\textit{dotted}) and total leakage (\textit{dot-dashed}) are calculated for the large value of $\beta=0.2$ to exaggerate the change due to the motion of the cluster. The observed quadrupole by the cluster (\textit{solid red}) has a different frequency function compared to the primordial quadrupole. Notice that the Doppler and aberration effects cancel each other at 86 GHz for this particular value of $\betahat$, so the total leakage vanishes at this frequency.} 
 	\label{fig:leakage}
 \end{figure}

 \begin{multline}\label{Ylm_integrand_expand}
 Y_{\ell m}(\gammahat_{cmb})=Y_{\ell m}(\gammahat_c)\\
 +\beta\sum\nolimits_{j,M'}^{1,\ell+j}  \Omega^{(1,j)}_{\ell m M'}(\betahat) Y_{(\ell+j)M'}(\gammahat_c)  \\
 +\frac{\beta^2}{2} \sum\nolimits_{j,M'}^{2,\ell+j} \Omega^{(2,j)}_{\ell mM'}(\betahat)Y_{(\ell+j)M'}(\gammahat_c)+ O(\beta^3).
 \end{multline}
 The rotation operators $\Omega^{(k,j)}_{\ell m M'}(\betahat)$ consist of an active and passive rotation on spherical harmonics, performed by two Wigner-D matrices  (equation \eqref{Omega_def}). The operators $\Omega^{(1,0)}_{\ell m M'}(\betahat)$, $\Omega^{(2,+1)}_{\ell m M'}(\betahat)$ and $\Omega^{(2,-1)}_{\ell m M'}(\betahat)$ are identically zero. Unlike Doppler effect, the aberration expansion is not frequency dependent and it only induces an angular distortion to the multipoles. The sum limits of the index $j$ in equation \eqref{Ylm_integrand_expand} show that up to order $\beta^k$, aberration distorts the multipole $\ell$ by drawing in its $k^\text{th}$ nearest neighbors.
 
 In a moving frame, the Doppler and aberration effects both induce geometrical leakage of  nearby multipoles into each other, but since their dependence on $\betahat$ is not the same, their contribution to the final observed $a^{I_c}_{\ell m}$ can be different. Since the Doppler effect is related to the frequency derivative of the anisotropies, it is usually the dominant effect (appendix \ref{sec:appA}). The derivatives can amplify the Doppler terms by large factors and also shift the peak of the frequency spectrum of the $a^{I_c}_{\ell m}$s observed by the cluster.

\subsection{\label{sec:IIIB}The Observed Multipoles in the Cluster's Frame}\label{The Observed Multipoles subsection}
After substituting equations \eqref{Doppler_expand} and \eqref{Ylm_integrand_expand} in \eqref{alm_cluster}, we can easily integrate over $\gammahat_c$ and obtain the following expression for the observed $a^{I_c}_{\ell m}$s in the cluster's frame:

\begin{multline}\label{alm_final}
a^{I_c}_{\ell'm'}(\nu_c)=a^{I}_{\ell'm'}(\nu_c) +\beta^2\Dnu^{(20)}a^{I}_{\ell'm'}(\nu)\\  +\sum_{p+q \in \{1,2\}}\sum_{\ell,m}\beta^{p+q}\Gij{p}{q}_{\ell' m'}^{\ell m} (\betahat) \Dnu^{(pp)}a^{I}_{\ell m}(\nu)+O(\beta^3).
\end{multline}
The geometrical kernel $\Gij{p}{q}^{\ell m}_{\ell'm'}(\betahat)$ is a modified version of $Y^*_{N n}(\betahat)$ and $\Omega^{(k,j)}_{\ell m M'}(\betahat)$ in equations \eqref{Doppler_expand} and \eqref{Ylm_integrand_expand} and it shows the geometrical leakage of order $\beta^p$ in Doppler effect and order $\beta^q$ in aberration, of the multipole $\ell$ to the observed multipole $\ell'$. For example, $\Gij{1}{0}^{3m}_{2m'}$ shows the geometrical leakage of the octupole into the observed quadrupole, entirely due to the Doppler effect up to first order in $\beta$ (see appendix \ref{sec:appB1c} equations \eqref{Gij}-\eqref{Gij2} for a  list of the geometrical factors).

\begin{figure}\label{new_quad}
	\centering
	\includegraphics[width=1.05\linewidth]{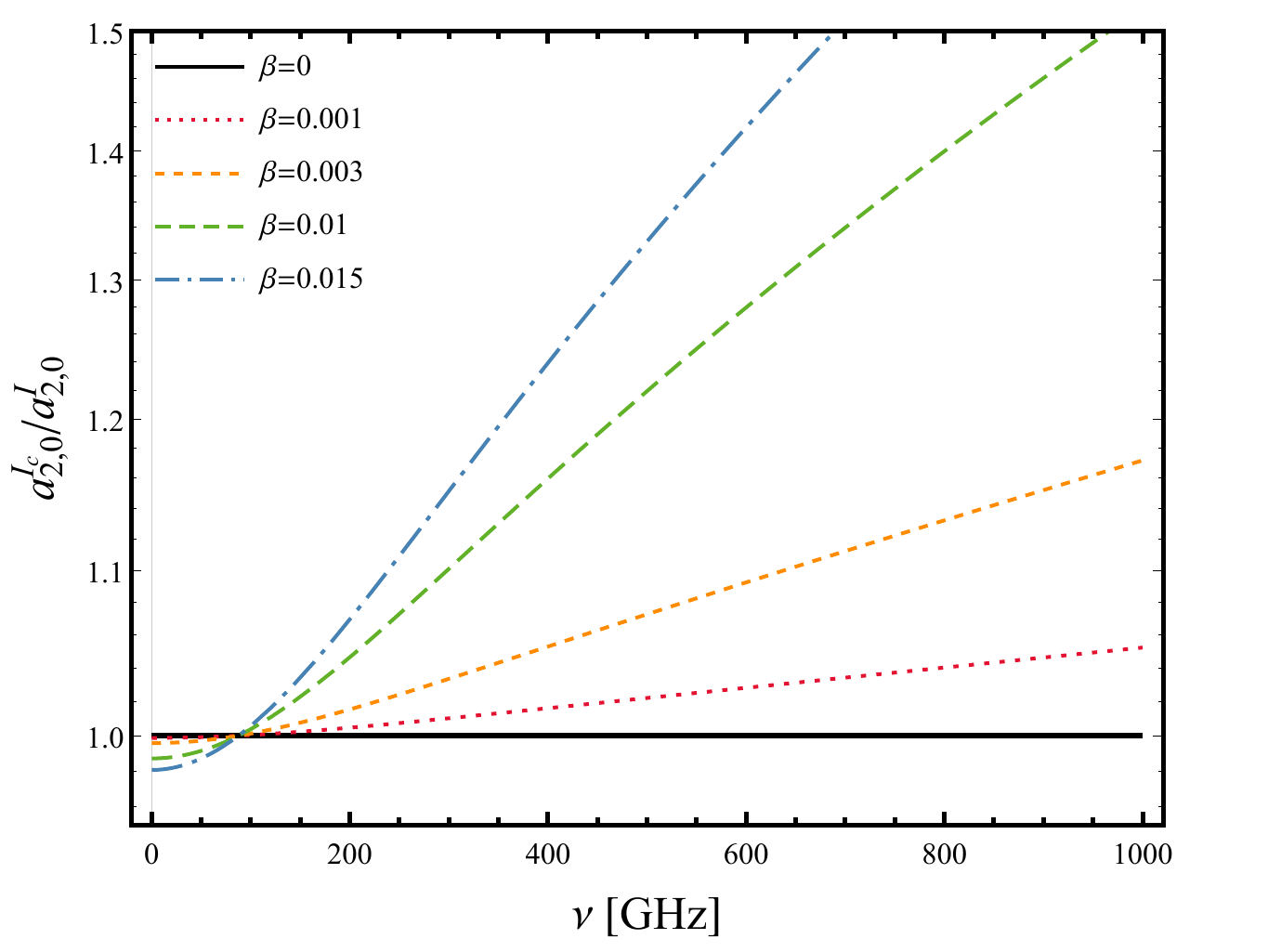}
	\caption{The ratio of the $a^{I_c}_{2,0}(\nu)$ coefficient of the quadrupole observed by a cluster moving in the $-\hat{\bm z}$ direction, and the primordial $a^{I}_{2,0}(\nu)$ observed in the CMB rest frame. Higher values of $\beta$ increase the leakage of the octupole and the dipole and amplify the quadrupole by 5\%-20\% in the frequency range of 200-400 GHz. For a typical velocity of 1000 km/s this change is about 2\%-10\% in the 200-600 GHz range.}
	\label{fig:newquad}
\end{figure}

Similarly to what was done in \S\ref{sec:IIB2}, assuming that the CMB has a pure blackbody spectrum in its rest frame, the frequency dependence of the $a^{I}_{\ell m}(\nu)$ coefficients can be easily separated from the spatial part using equations \eqref{alm(I)_to_alm(T)}. For convenience, we introduce the following frequency functions

  \begin{subequations}
  	\begin{align}
  	\tilde{B}^{(ij)}_{\nu_c}(\Tz) & \equiv \Dnu^{(ij)} \tilde{B}_{\nu}(\Tz),\\
  	\tilde{F}^{(ij)}_{\nu_c}(\Tz) & \equiv \Dnu^{(ij)} \tilde{F}_{\nu}(\Tz).
  	\end{align}
  \end{subequations}
After substituting these functions in equation \eqref{alm_final} and writing the RHS in terms of the thermodynamic temperature coefficients $a^{T_z}_{\ell m}$, we can easily find the leakage of the primordial anisotropies into each multipole observed by the moving cluster. For example, figure \ref{fig:leakage} shows the frequency functions of the leakage of the dipole and octupole into the observed $a^{I_c}_{2,0}$ quadrupole coefficient by a cluster that is moving with $\vec{\bm \beta}=-0.2\hat{\bm z}$. The large value of the peculiar velocity is chosen to exaggerate the total effect. Here the octupole coefficients are also adopted from Ref. \citep{Polastri:2015rda} but the dipole coefficients are simulated. It is evident from figure \ref{fig:leakage} that the Doppler and aberration leakages have different frequency functions and for the particular value of $\betahat=-\hat{\bm z}$, the two effects cancel each other at 86 GHz. In the plot, the apparent cancellation between the nonmoving quadrupole, and the aberration leakage of the dipole and octupole is just a coincidence due to the choice of $\beta=0.2$. This value is roughly equal to the ratio of the local $a^{T}_{2,0}$ coefficient to the amplitude of the dipole and octupole moments used in the plot.

Figure \ref{fig:newquad} shows the relative change in the observed $a^{I_c}_{2,0}$ coefficient for peculiar velocities ranging between $\beta=0.001~ (v=300 \text{km/s})$ to $\beta=0.015~ (v=4500 \text{km/s})$. As we can see in this figure, the frequency at which the Doppler and aberration effects cancel each other does not change for different values of $\beta$. For $\nu>86$ GHz the Doppler leakage becomes dominant with a larger frequency weight compared to the aberration leakage. As a result, the change in the quadrupole observed by the cluster at these frequencies can be relatively large, despite the small values of $\beta$.

It is worth mentioning that equation \eqref{alm_final} can be also used in the context of deboosting/deaberrating the CMB multipoles in the local frame. Writing this equation in the local frame allows us to easily find the observed multipoles (LHS) in terms of the primordial ones (RHS). The most important feature of this equation is that it shows the dependence of the overall frequency function on the direction of motion of the frame $\betahat$. This dependence is not clearly recognizable in the usual calculations of the aberration kernel, due to the convenient choice of $\betahat=\hat{\bm z}$. In that case one can in principle still recover the same results by applying appropriate rotations to the $a^{I}_{\ell m'}$s numerically. In fact the geometrical factors $\Gij{p}{q}^{\ell m}_{\ell'm'}(\betahat)$ in equation \eqref{alm_final} can be interpreted as these ``appropriate rotations" applied to the $a^{I}_{\ell m}$s. In appendix \ref{sec:appBII}, equation \eqref{alm_final} is explicitly calculated for the first few multipoles of the CMB in a boosted frame. 

 In the following section we calculate the polarization and intensity induced by the $a^{I_c}_{2m'}$ coefficients in the direction of a moving cluster, and their dependence on the primordial dipole and octupole, which will be reflected through the cluster in the TinIn and TinPol effects.

\subsection{\label{sec:IIIC}Induced Signals in the Direction of the Cluster}
\subsubsection{\label{sec:IIIC1}TinPol and kSZ polarization effects}
Using the transformation between the Intensity multipoles in the CMB frame and the cluster's frame, it is now easy to find the Intensity and polarization induced in the direction of a moving cluster. We simply need to replace the terms on the RHS of equations \eqref{TinIn} and \eqref{TinPol} with those calculated from equation \eqref{alm_final}. Although the results derived here will be only valid for an observer in the cluster's frame, converting it to any arbitrary frame would be trivial using the transformations \eqref{doppler_freq} and \eqref{aberration}. One can first transform the signal back to the CMB rest frame and then Lorentz boost it into the frame of a general observer using the same equations. This procedure is outlined in great detail in Refs. \citep{Chluba2012c} and \citep{Nozawa:2013rna}.

We start with the change in polarization, because unlike intensity it directly probes the quadrupole observed by the cluster, and it is not distorted by the higher multipoles of the temperature anisotropies that induce the bSZ effect (see equations \eqref{Delta_I_gamma} and \eqref{Delta_Q+iU_gamma} in \S \ref{sec:IIB1}). Equation \eqref{TinPol}
in the cluster's frame is

\begin{equation}\label{Delta_Q+iU_cluster}
\frac{\Delta (Q_{\nu_c}\pm iU_{\nu_c})(\gammahat_c) }{\Delta\tau} =-\delta (Q_{\nu_c}\pm iU_{\nu_c})^{(2)}(\gammahat_c) \\
\end{equation}
where the quadrupole-induced polarization on the RHS is now proportional to the primordial temperature quadrupole, as well as its first neighboring multipoles (the dipole and the octupole) to first order in $\beta$, and its second neighbors (the monopole and the hexadecapole) to second order in $\beta$ and so on. The first few terms of this expansion are as follows

\begin{align}\label{newTinPol}
\delta (Q_{\nu_c}\pm iU_{\nu_c})^{(2)}(\gammahat_c)=& \text{T}_2\text{inPol}+\text{kSZPol}\\
+&\text{T}_1\text{inPol}+\text{T}_3\text{inPol}+O(\beta^2),\nonumber
\end{align}
with the following definitions

 \begin{figure}[t]\label{newtinPol}
 	\centering
 	\includegraphics[width=1.0\linewidth]{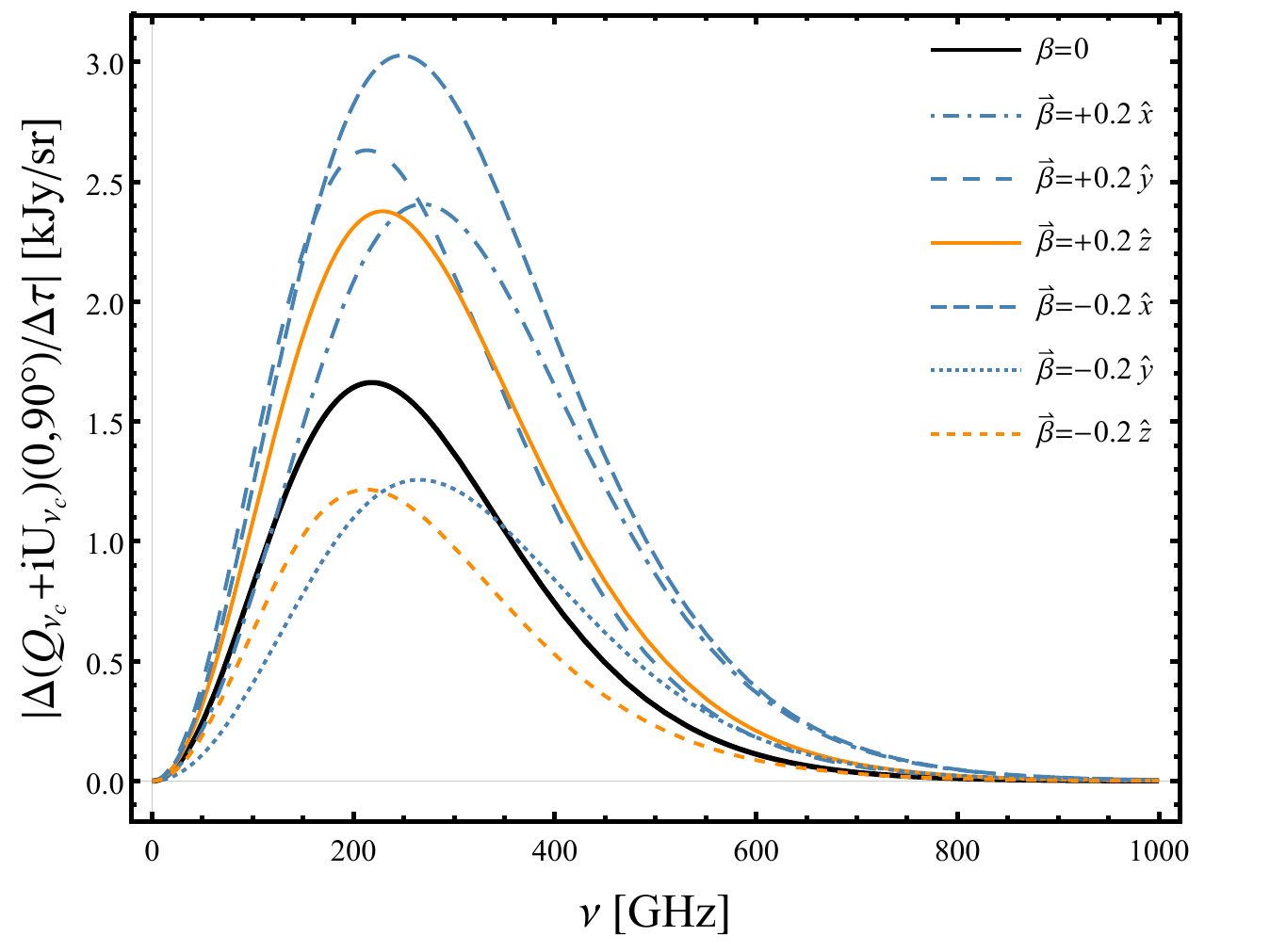}
 	\medskip
 	\par \medskip
 	\includegraphics[width=1.0\linewidth]{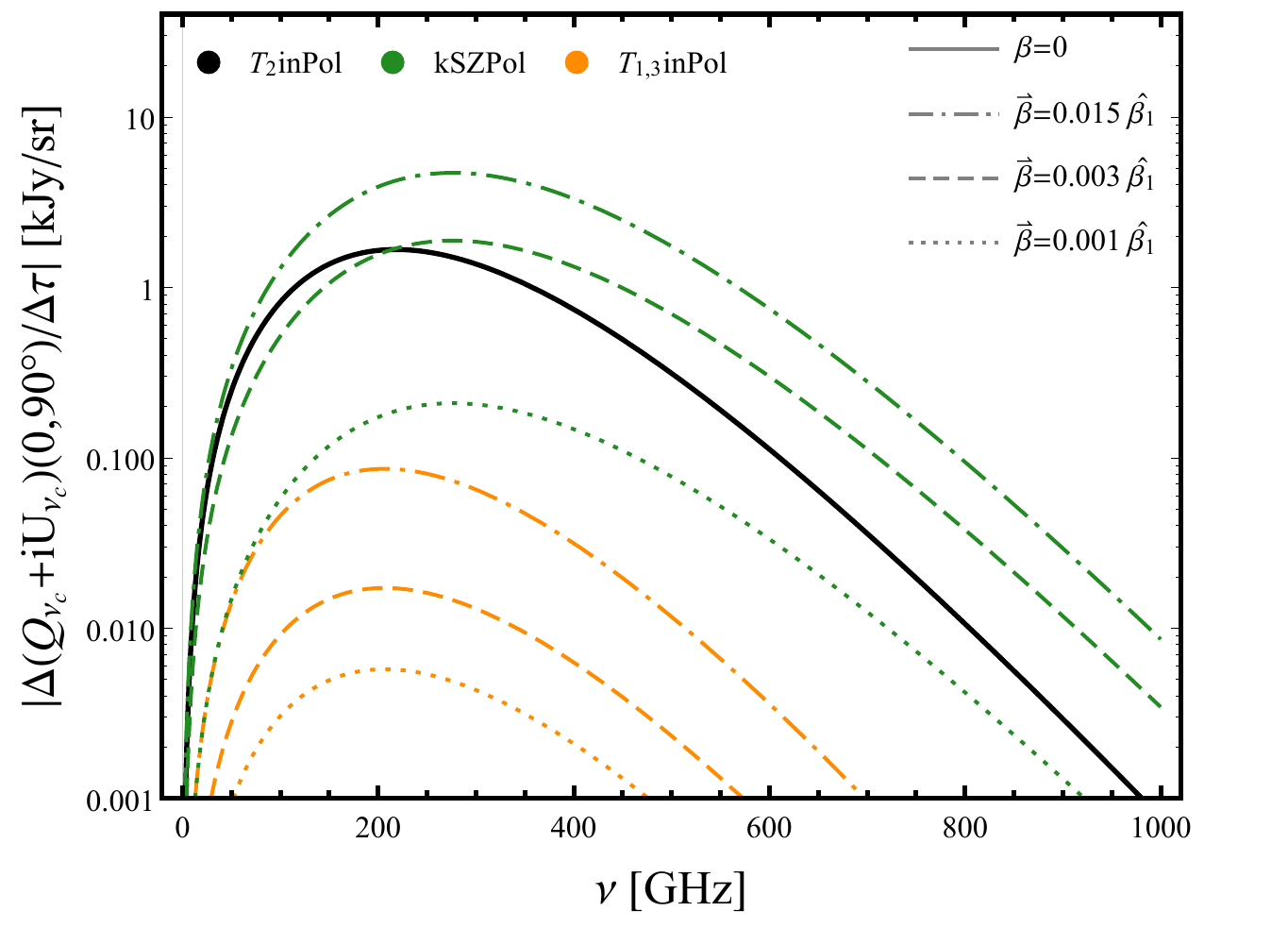}
 	\caption{(\textit{top}) Temperature quadrupole-induced Polarization  (T$_{2}$inPol)  effect (\textit{solid black}) and its modifications due to the dipole and octupole in the $(l,b)=(0^\circ,90^\circ)$ direction for different values of $\betahat$.  The large value of $\beta=0.2$ is chosen to exaggerate the differences between the frequency functions. Since the ratio between Doppler and aberration leakage of the dipole and octupole depends on the direction of motion of the cluster, the frequency function of the total signal changes with $\betahat$. The \textit{orange} and \textit{blue} colors correspond to clusters with only radial and transverse velocity vectors. (\textit{bottom}) Induced polarization by the quadrupole moment in comparison with that of the dipole and octupole  (T$_{1,3}$inPol) and  kSZ Polarization (kSZPol) for clusters moving in the $\betahat_1=(-54^\circ,27^\circ)$ direction with 4500 km/s (\textit{dot-dashed}), 900 km/s (\textit{dashed}) and 300 km/s (\textit{dashed}). The $\beta=0.015$ kSZPol signal is scaled by a factor 0.1. }
 	\label{fig:newtinPol}
 \end{figure}

\begin{align}\label{T2inPol}
\text{T}_2\text{inPol}\equiv
\frac{\sqrt{6}}{10}\Fnuc(\Tz)\sum\nolimits_{m}^{2} ~_{\mp 2}Y_{2m}(\gammahat_c)a^{T_z}_{2m},\hspace{4em}
\end{align}
\begin{multline}\label{kSZPol}
\text{kSZPol}\equiv
\frac{\sqrt{6}}{10}\beta^2\Bnuc^{(22)}(\Tz)\sum\nolimits_{m',m}^{2,2} \Gij{2}{0}^{00}_{2m'}(\betahat)\\\times_{\mp 2}Y_{2m'}(\gammahat_c)a^{T_z}_{00},
\end{multline}
\begin{multline}\label{T1inPol}
\text{T}_1\text{inPol}\equiv\\
\frac{\sqrt{6}}{10}\beta\sum\nolimits_{m',m}^{2,1}\left(\Gij{1}{0}^{1m}_{2m'}(\betahat)\Fnuc^{(11)}(\Tz)
+\Gij{0}{1}^{1m}_{2m'}(\betahat)\Fnuc(\Tz)\right)\\
\hspace{13em}\times~_{\mp 2}Y_{2m'}(\gammahat_c)a^{T_z}_{1m},
\end{multline}
\begin{multline}\label{T3inPol}
\text{T}_3\text{inPol}\equiv\\
\frac{\sqrt{6}}{10}\beta\sum\nolimits_{m',m}^{2,3}\left(\Gij{1}{0}^{3m}_{2m'}(\betahat)\Fnuc^{(11)}(\Tz)
+\Gij{0}{1}^{3m}_{2m'}(\betahat)\Fnuc(\Tz)\right)\\
\hspace{13em}\times~_{\mp 2} Y_{2m'}(\gammahat_c)a^{T_z}_{3m}.
\end{multline}
 The first term in equation \eqref{newTinPol} is the quadrupole-induced polarization (T$_2$inPol) that was already present for a non-moving cluster (compare with equation \eqref{TinPol}). The monopole-induced term, defined in equation \eqref{kSZPol}, is the well known kSZ polarization effect (kSZPol). In order to make this expression look more familiar, we write the frequency function and geometrical factor explicitly to obtain

 \begin{multline}\label{kSZPol2}
 \frac{\Delta (Q_{\nu_c}\pm iU_{\nu_c})^{\text{kSZ}}(\gammahat_c) }{\Delta\tau} =
 \frac{\beta^2}{20}\sin^2\vartheta_\beta ~e^{\pm 2i\varphi_\beta}\\\times \Fnuc(\Tz)x_c\coth(x_c/2),
 \end{multline}
where $x_c\equiv h\nu_c/k\Tz$ and $\vartheta_\beta$ and $\varphi_\beta$ are the polar and azimuthal Euler angles between $\gammahat_c$ and $\betahat$ defined as  \citep{Okamoto:2002ik}
\begin{multline}
\cos\vartheta_\beta=\cos\theta_c \cos\theta_\beta+\sin\theta_c\sin\theta_\beta \cos(\phi_c-\phi_\beta),\\
\cot\varphi_\beta=\cos\theta_\beta\cot(\phi_c-\phi_\beta)-\cot\theta_c\sin\theta_\beta\csc(\phi_c-\phi_\beta).
\end{multline}
In equation \eqref{kSZPol2}, $\beta \sin\vartheta_\beta=\beta_\perp$ is the transverse component of the peculiar velocity with respect to the observer's line of sight. This expressions generalizes the the results of \citep{Sunyaev1981,Audit1999} in agreement with \citep{Sazonov1999}. The $e^{\pm 2i\varphi_\beta}$ factor decomposes the polarization signal into the $Q$ and $U$ components, respectively with factors of $\cos(2\varphi_\beta)$ and $ \sin(2\varphi_\beta)$.  Therefore the combination of the $Q$ and $U$ polarization maps can be used to determine the exact three dimensional direction of the cluster's velocity vector.

\begin{figure}[t]
	\centering
	\text{217 GHz}\par \medskip
	\includegraphics[width=1.0\linewidth]{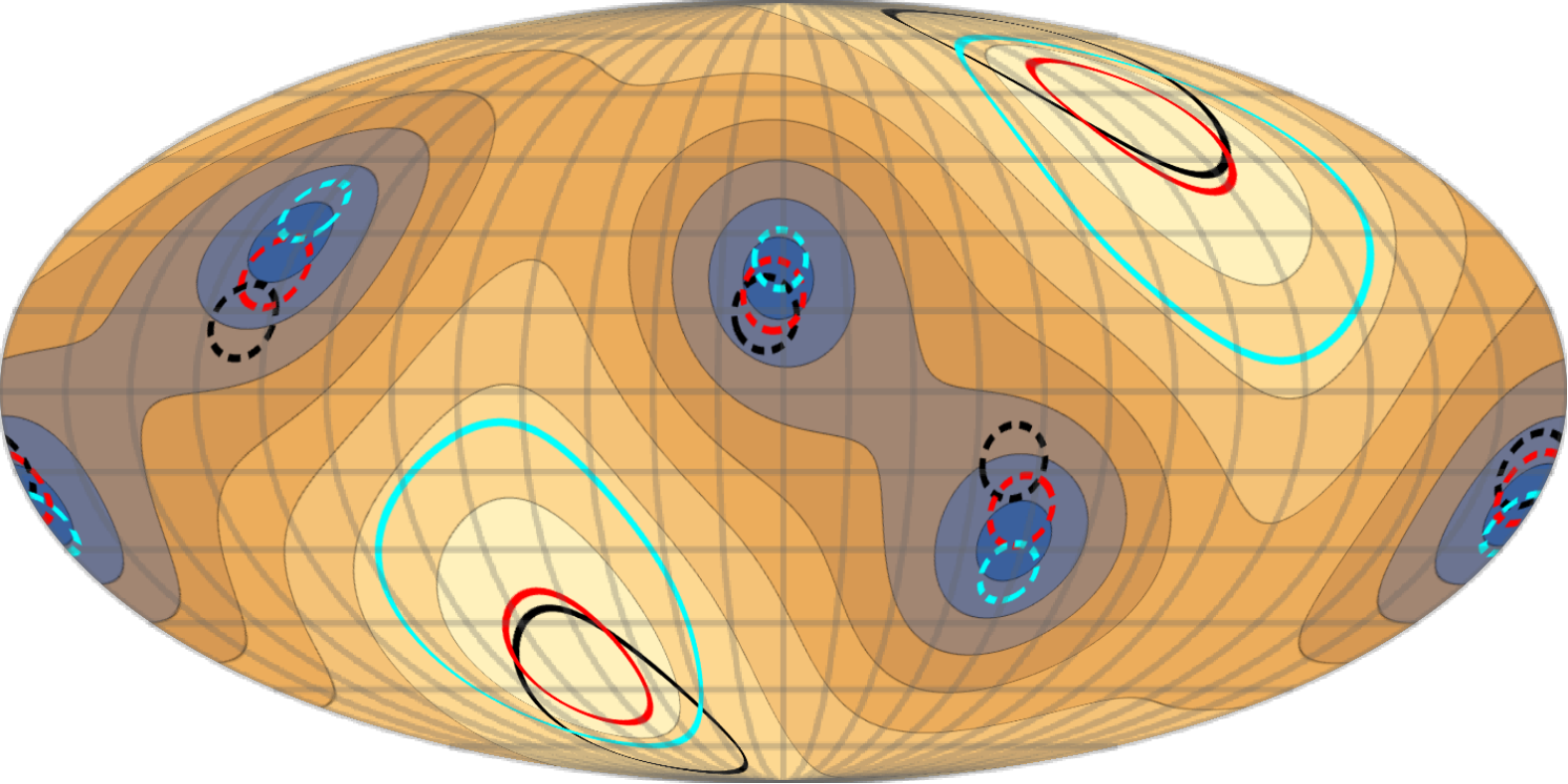}
	\includegraphics[width=1.0\linewidth]{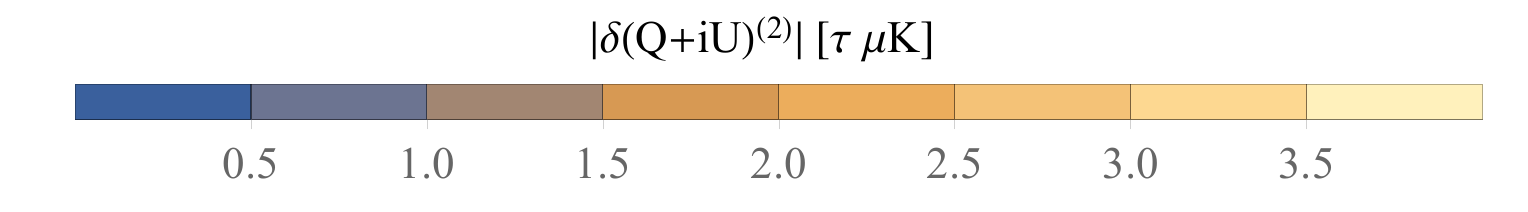}
	\caption{ Temperature-induced polarization for clusters moving with the peculiar velocity $\beta=0.05$ in the $-\hat{\bm z}$ direction at 217 GHz. The \textit{dashed} circles depict the 0.5 $\tau \muK$  contour lines at 300 GHz (\textit{cyan}) and 135 GHz (\textit{red}) and for non-moving clusters at 217GHz  (\textit{black}). Similarly the \textit{solid} lines show the 3.5 $\tau \muK$ contour lines over different frequency channels. The shift in the location of the maxima (minima) of the signal between 217 GHz and 300 GHz channels is $23^\circ$ ($11^\circ$ and $22^\circ$).  }
	\label{fig:TIP-moving-217GHz}
\end{figure}

The second line of equation \eqref{newTinPol} shows the dipole and octupole-induced polarization (T$_{1}$inPol and T$_{3}$inPol). We will refer to these terms collectively as T$_{1,3}$inPol. The most interesting feature of these new terms is that they induce different frequency and angular distortions to the TinPol at different frequency bands. Note that in equations \eqref{T2inPol}-\eqref{T3inPol}, the dependence of the signals on the observation direction $\gammahat_c$, and the velocity direction of the cluster $\betahat$ are separated as a result of our generalized approach. For the T$_{1,3}$inPol effect, changing any of these directions changes the total frequency function of the signal. In figures \ref{fig:newtinPol} and \ref{fig:TIP-moving-217GHz}, by fixing one of these directions, we show the dependence of the signal on the other one. The top panel of figure \ref{fig:newtinPol} shows how much the dipole and octupole can change the TinPol signal and its frequency dependence in the direction of the north galactic pole for an exaggerated value of $\beta$. The T$_2$inPol effect, which is the only term that does not depend on the velocity of the cluster, reaches its maximum at 218 GHz. The peak value of this signal is 1.7 $\tau$kJy/sr $(3.6 \tau\muK)$ which for a cluster with optical depth $\tau\approx0.02$ is equal to 34 Jy/sr $(0.07 \muK)$. In the bottom panel we have separated the T$_{1,3}$inPol for comparison with the T$_2$inPol  and kSZPol for realistic values of $\beta$. Since the leakage due to the Doppler effect is typically dominant over the aberration effect (see figure \ref{fig:leakage}), it shifts the position of the peak of the dipole and octupole-induced polarizations with respect to the quadrupole-induced signal. The direction of motion of the cluster $\betahat$ is chosen to maximize the T$_{1,3}$inPol in this specific direction in the sky. For bullet-like clusters ($\beta=0.015$), kSZPol is dominant over the T$_2$inPol but for low velocities ($\beta=0.001$) it is approximately smaller by a factor of 10. The reason for kSZPol's fast drop is its dependence on $\beta_\perp^2$. Although T$_{1,3}$inPol is much smaller than kSZPol at high velocities, it induces a 10\% correction to T$_2$inPol at 217 GHz. For smaller velocities this effect is about 5\% of the kSZPol at low frequencies. Although kSZPol is typically the dominant effect in equation \eqref{newTinPol} for larger values of $\beta$, it is nevertheless distinguishable from the other TinPol effects due to its distinct frequency dependence and large amplitude. The leakage of the hexadecapole has been neglected in equation \eqref{newTinPol} because it only contributes 0.2\% (2\%) to the total polarization in the Rayleigh-Jeans (Wien) frequencies.  

\begin{figure}
	\centering
	\text{300 GHz}\par \medskip
	\includegraphics[width=1.0\linewidth]{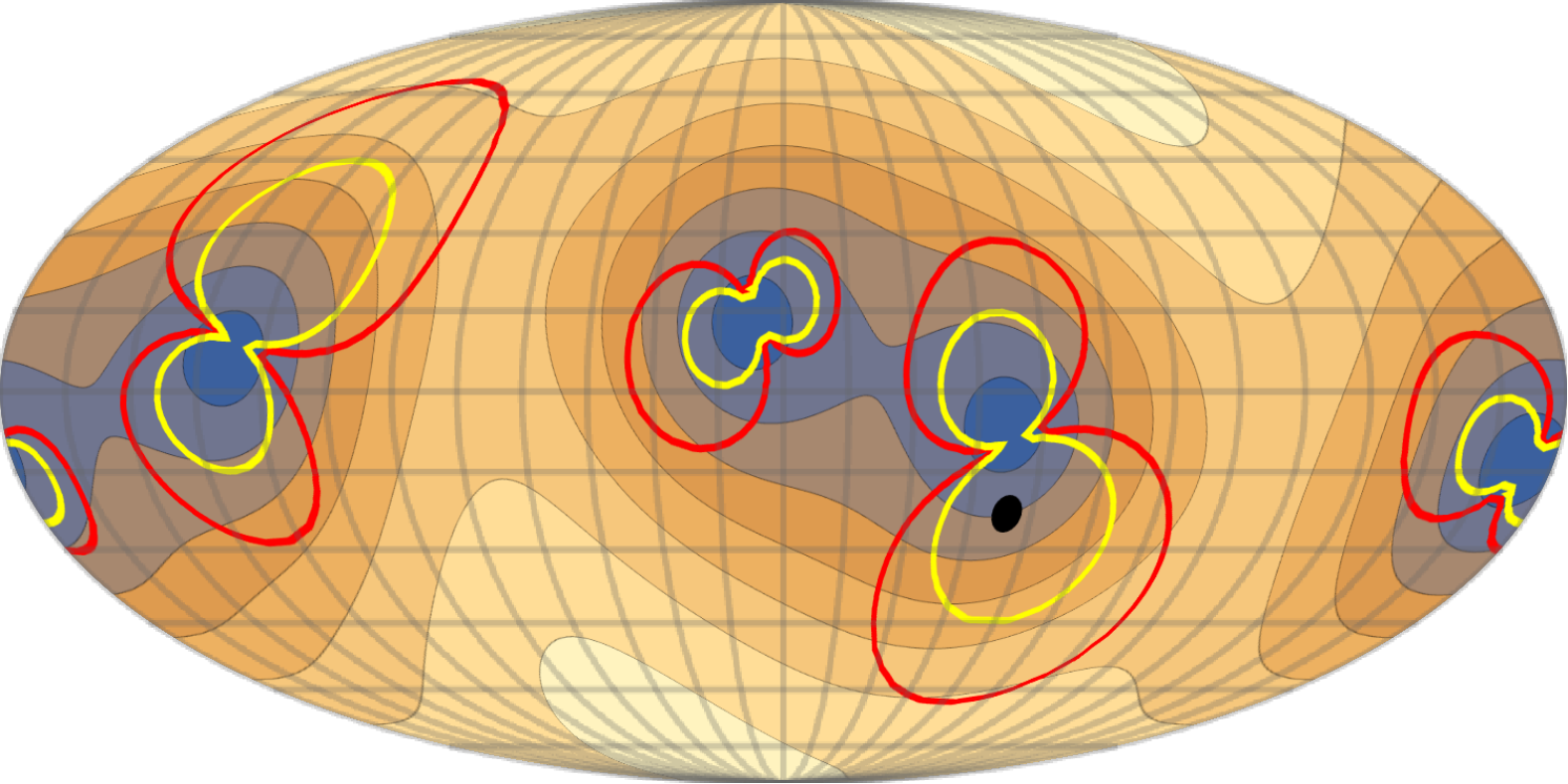}
	\includegraphics[width=1.0\linewidth]{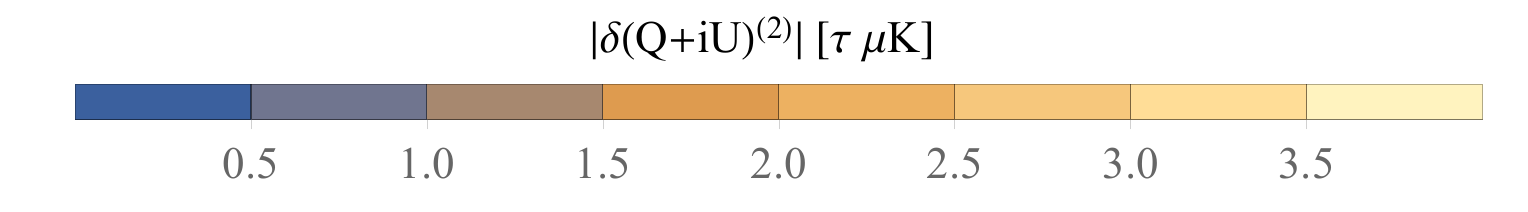}
	\par \medskip
	\includegraphics[width=1.0\linewidth]{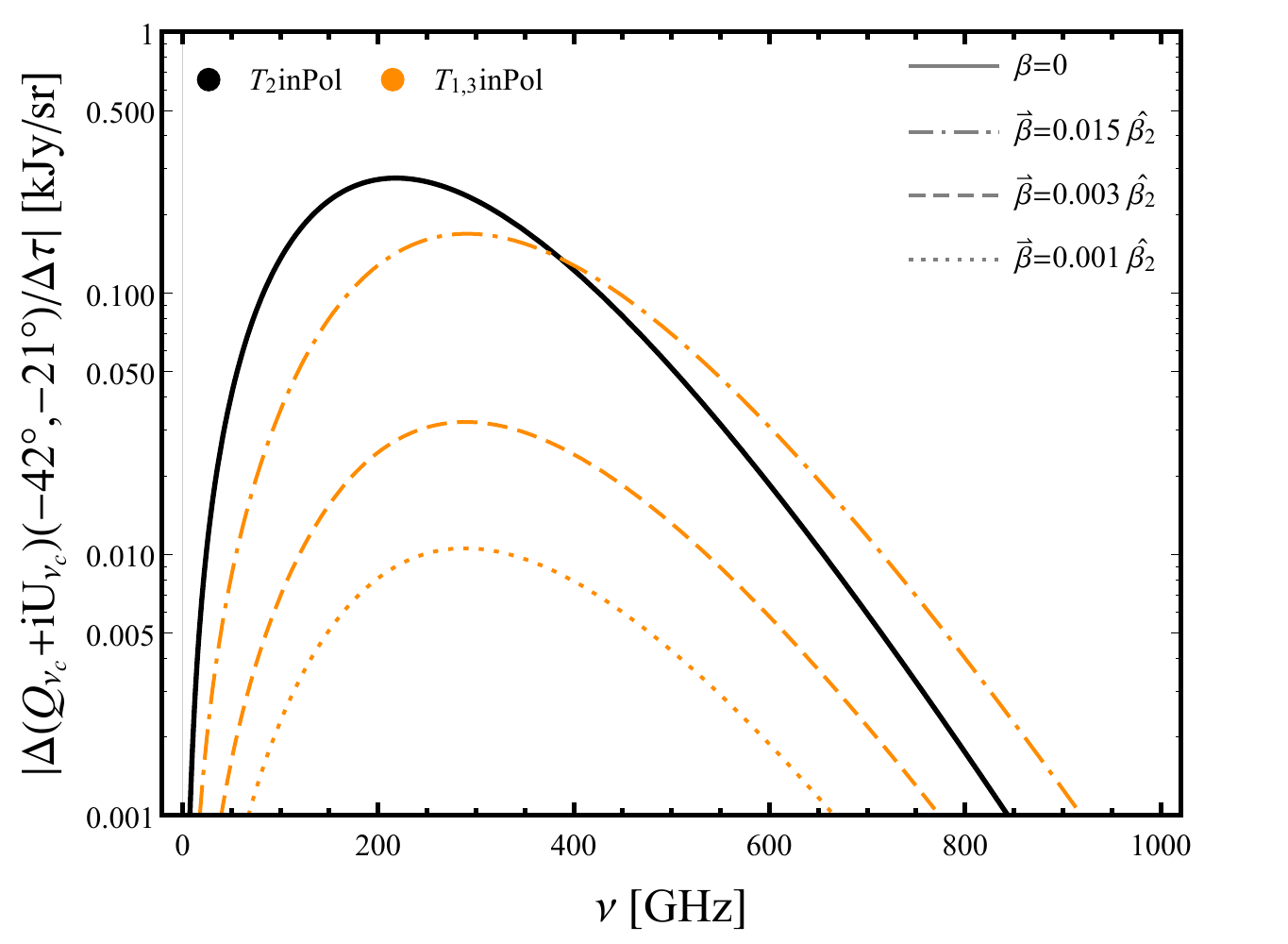}
	\caption{(\textit{top}) Mollweide projection of the TinPol effect for clusters moving with the peculiar velocity $\beta=0.015$ in the $\betahat_2=(-135^\circ,45^\circ)$ direction. This direction is chosen to maximize the T$_{1,3}$inPol signal. The \textit{red} (\emph{yellow}) contours show the ares within which the T$_{1,3}$inPol is larger than 10\% (20\%) of the T$_2$inPol signal. (\textit{bottom}) Polarization signals in the direction of the black dot in the top panel. In this direction T$_{1,3}$inPol becomes larger than T$_2$inPol above 420 GHz for $\beta=0.015$ (\textit{dot-dashed}).  At 289 GHz the peak of the T$_{1,3}$inPol is respectively 70\%, 13\% and 4\% of the T$_{2}$inPol for 4500 km/s (\textit{dot-dashed}), 900 km/s (\textit{dashed}) and 300 km/s (\textit{dashed}).}
	\label{fig:TIP-moving-300GHz}
\end{figure}

Doppler and aberration leakages of the dipole and octupole also induce an angular distortion to the TinPol signal. Figure \ref{fig:TIP-moving-217GHz} shows the shift in the location of the minima and maxima of the signal to represent the angular change of the overall signal at different frequencies, for clusters moving with $\vec{\bm \beta}=-0.05 \hat{\bm{z}}$. The maxima and minima of the TinPol signal are respectively $32^\circ$ and $14^\circ$ apart at 217 GHz and 300 GHz. For a more realistic velocity $\beta=0.015$, these angle separations are about $15^\circ$ for the maxima and $5^\circ$ for the minima. Note that the angular distortions in this plot are calculated for one particular direction of motion of clusters. Figure \ref{fig:TIP-moving-217GHz} simply shows that the induced angular distortions are frequency dependent and unless the clusters in some observation direction at a certain redshift bin have a coherent bulk motion, the expected signal would not change as described in this plots. Indeed, for a more realistic map one would need to populate the sky with clusters moving in different directions, but even in that case the angular dependence of the signal will change at different frequencies. This feature can be extremely helpful in a multi-frequency and whole-sky survey to identify and extract the TinPol effects from foreground distortions and even kSZ and tSZ effects which are not expected to have similar distributions over the sky.  Also, the angle dependence of this distortion, changes the ratio between T$_{1,3}$inPol and T$_2$inPol effects at different observation directions, and obviously makes T$_{1,3}$inPol the dominant effect in the areas where T$_2$inPol is small. 
\begin{figure}[t!]
	\centering
	\includegraphics[width=1.05\linewidth]{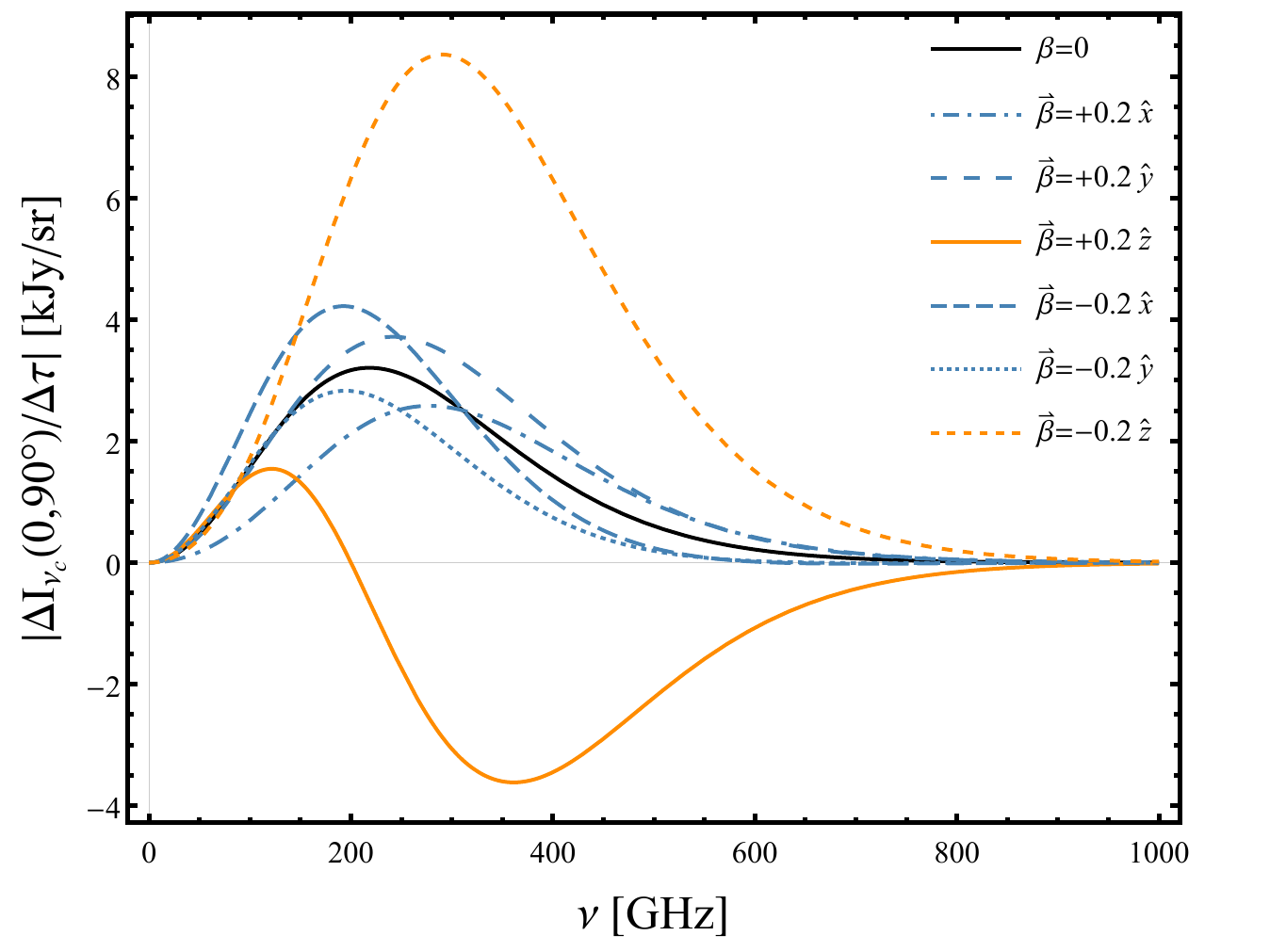}
	\medskip
	\par \medskip
	\includegraphics[width=1.05\linewidth]{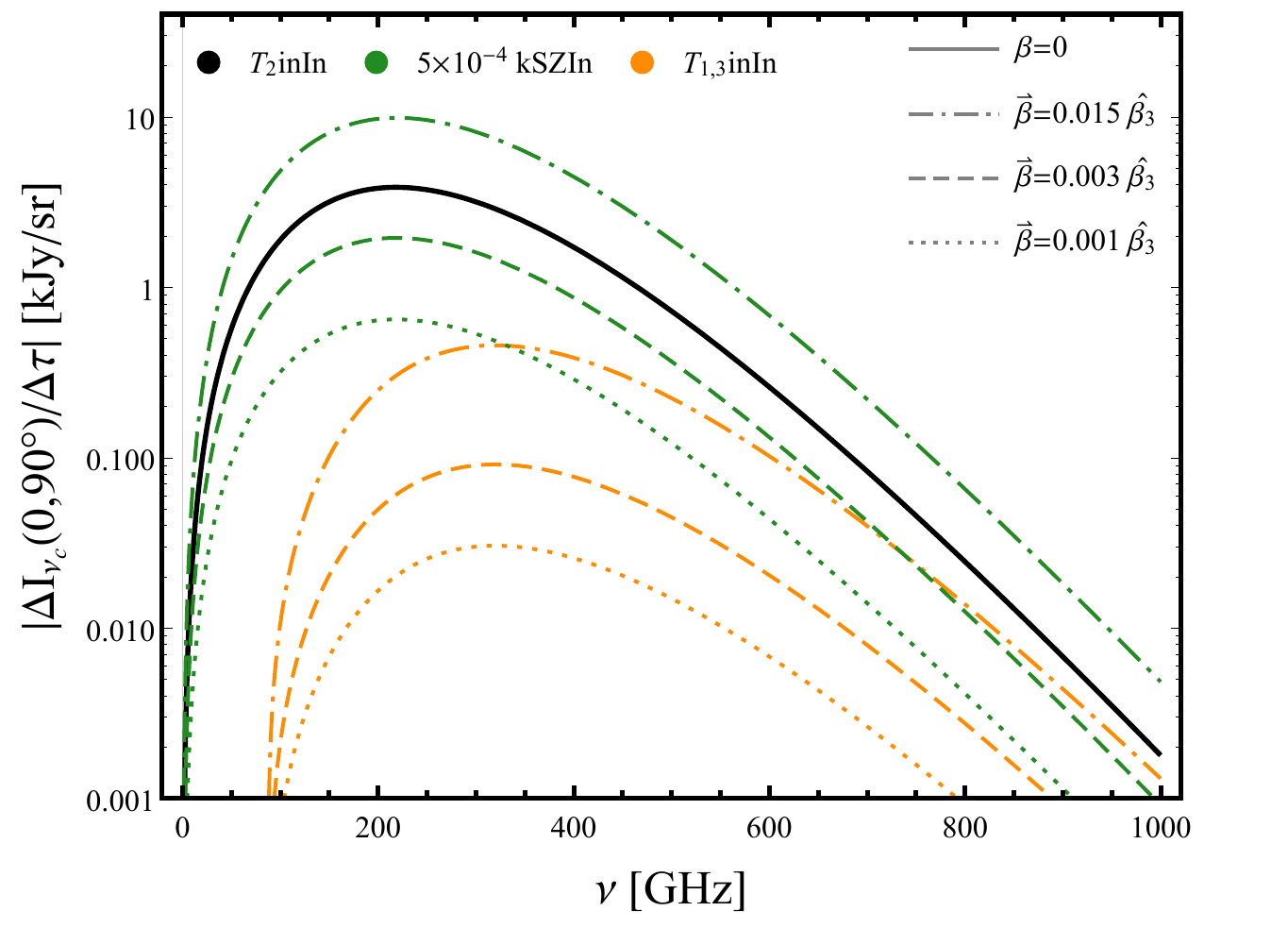}
	\caption{(\textit{top})	Temperature quadrupole-induced intensity (T$_{2}$inIn) effect (solid black) and its modifications due to the dipole and octupole in the $(l,b)=(0^\circ,90^\circ)$ direction for different values of $\betahat$. The maximal change is in the direction of clusters moving with a radial velocity. (\textit{bottom}) Induced intensity by the quadrupole moment in comparison with that of the dipole and octupole (T$_{1,3}$inIn) and  the kSZ intensity (kSZIn) effect rescaled by $5\times10^{-4}$ for clusters moving in the $\betahat_3=(-94^\circ,76^\circ)$ direction with 4500 km/s (\textit{dot-dashed}), 900 km/s (\textit{dashed}) and 300 km/s (\textit{dashed}). 
	}
	\label{fig:newtinIn}
\end{figure}

The top panel of figure \ref{fig:TIP-moving-300GHz} shows a similar map to figure \ref{fig:TIP-moving-217GHz} at 300GHz, which is the frequency at which the peak of T$_{1,3}$inPol typically lies. We can see that T$_{1,3}$inPol is larger than 10\% (20\%) of T$_2$inPol over roughly 20\% (10\%) of the sky. Here, the average contribution of the T$_{1,3}$inPol over the whole sky is 10\%. The bottom panel shows different polarization signals in an arbitrary direction $(l,b)=(-54^\circ,-27^\circ)$ which is indicated with a black dot in the top panel. In this direction the amplitude of T$_{1,3}$inPol for $\beta=0.015$ is comparable to T$_2$inPol and even becomes dominant at 386 GHz. The peak value of T$_{1,3}$inPol is at 289 GHz and at this frequency it is as large as 70\% of the T$_{2}$inPol. For the smaller values of $\beta=0.003$ and $\beta=0.001$, T$_{1,3}$inPol is respectively 13\% and 4\% of the T$_{2}$inPol at the peak frequency. 

\subsubsection{\label{sec:IIIC2}TinIn and kSZ intensity effects}

Similar to the case of polarization, for intensity we can rewrite equation \eqref{TinIn} in the cluster's frame as 
\begin{equation}\label{Delta_I_cluster}
\frac{\Delta I_{\nu_c}(\gammahat_c) }{\Delta\tau}  = -\delta I_{\nu_c}^{(1)}-\delta I_{\nu_c}^{(2)}-\sum_{\ell=3}^{\infty}\delta I_{\nu_c}^{(\ell)},\\
\end{equation}
where the new bSZ$^{(1)}$ in the cluster's frame is
\begin{multline}\label{newbSZ1}
\delta I_{\nu_c}^{(1)}= \text{T}_1\text{inbSZ}^{(1)}+\text{kSZIn}^{(1)}+\text{T}_2\text{inbSZ}^{(1)}+O(\beta^2)
\end{multline}
where 

\begin{equation}\label{T1inbSZ}
\text{T}_1\text{inbSZ}^{(1)}\equiv \sum\nolimits_{m}^{1}Y_{1m}(\gammahat_c)\Fnuc(\Tz)a^{T_z}_{1m}
\end{equation}
 is the dipole-induced bSZ term, which was already present even in the case of the non-moving cluster,
\begin{equation}\label{kSZIn1}
\text{kSZIn}^{(1)}\equiv \beta \sum\nolimits_{m'}^{1} \Gij{1}{0}^{00}_{1m'}(\betahat)\Bnuc^{(11)}(\Tz)~Y_{1m'}(\gammahat_c)a^{T_z}_{00},
\end{equation}
is the first order monopole-induced intensity, known as the kSZ intensity effect and

\begin{figure}[t!]
	\centering
	\text{217 GHz}\par \medskip
	\includegraphics[width=0.9\linewidth]{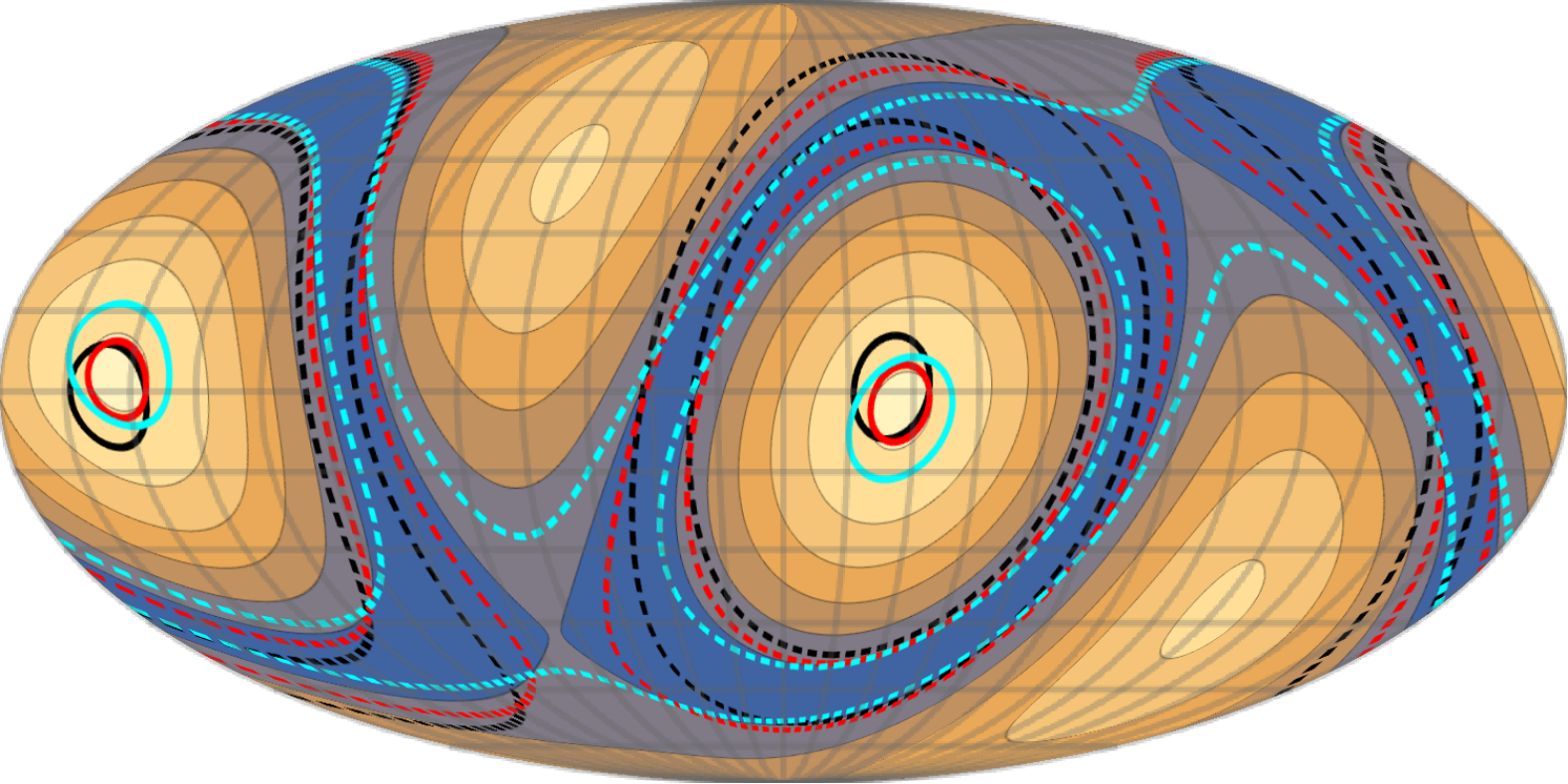}
	\includegraphics[width=0.9\linewidth]{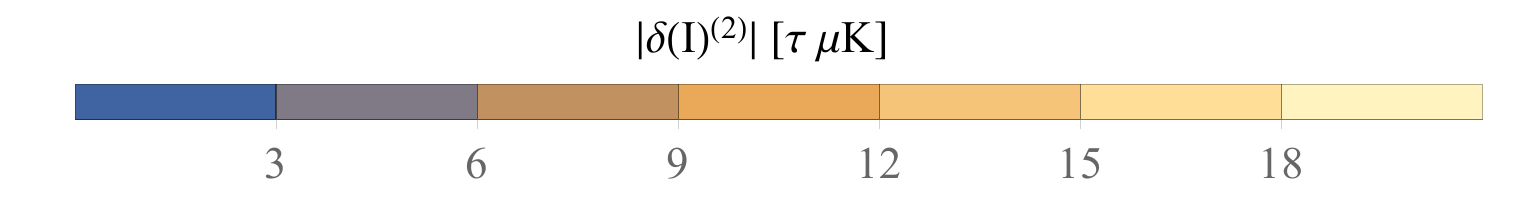}
	\caption{Temperature-induced intensity for clusters moving with the peculiar velocity $\beta=0.05$ in the $-\hat{\bm z}$ direction at 217 GHz. The \textit{dashed} circles depict the 3 $\tau \muK$  contour lines at 300 GHz (\textit{cyan}) and 135 GHz (\textit{red}) and at 217GHz for non-moving clusters (\textit{black}). The \textit{solid} lines show the 18 $\tau \muK$ contour lines over different frequency channels. Similar to the TinPol effect, the location of the maxima and minima of this signal change at different frequencies.}
	\label{fig:TIT-moving-217GHz}
\end{figure}

\begin{multline}\label{T2inbSZ}
\text{T}_2\text{inbSZ}^{(1)}\equiv \\
\beta \sum\nolimits_{m',m}^{1,2}\left(\Gij{1}{0}^{2m}_{1m'}(\betahat)\Fnu^{(11)}(\Tz)+\Gij{0}{1}^{2m}_{1m'}(\betahat)\Fnu(\Tz)\right)\\
\times Y_{1m'}(\gammahat_c)a^{T_z}_{2m}
\end{multline}
is the leakage of the quadrupole into the dipole-induced bSZ. Similarly, the new TinIn in the cluster's frame would be equal to 
\begin{multline}\label{newTinIn}
\delta I_{\nu_c}^{(2)}= \text{T}_2\text{inIn}+\text{kSZIn}^{(2)}+\text{T}_1\text{inIn}+\text{T}_3\text{inIn}+O(\beta^2)
\end{multline}
where again 
\begin{equation}\label{T2inIn}
\text{T}_2\text{inIn}\equiv \frac{9}{10}\sum\nolimits_{m}^{2} Y_{2m}(\gammahat_c)\Fnuc(T_0)a^{T}_{2m} 
\end{equation}
is the quadrupole-induced intensity which was already derived for non-moving clusters,
\begin{equation}\label{kSZIn2}
\text{kSZIn}^{(2)}\equiv \frac{9}{10} \beta^2 \sum\nolimits_{m'}^{2} \Gij{2}{0}^{00}_{2m'}(\betahat)\Bnuc^{(22)}(T_0) ~Y_{2m'}(\gammahat_c)a^T_{00} 
\end{equation}
is the second order monopole-induced intensity (first relativistic correction to kSZIn) and

\begin{multline}\label{T1inIn}
\text{T}_1\text{inIn}\equiv \\
\frac{9}{10}\beta\sum\nolimits_{m',m}^{2,1}\left(\Gij{1}{0}^{1m}_{2m'}(\betahat)\Fnuc^{(11)}(\Tz)
+\Gij{0}{1}^{1m}_{2m'}(\betahat)\Fnuc(\Tz)\right)\\
\times Y_{2m'}(\gammahat_c)a^{T_z}_{1m},
\end{multline}

\begin{multline}\label{T3inIn}
\text{T}_3\text{inIn}\equiv \\
\frac{9}{10}\beta\sum\nolimits_{m',m}^{2,3}\left(\Gij{1}{0}^{3m}_{2m'}(\betahat)\Fnuc^{(11)}(\Tz)
+\Gij{0}{1}^{3m}_{2m'}(\betahat)\Fnuc(\Tz)\right)\\
\times Y_{2m'}(\gammahat_c)a^{T_z}_{3m}
\end{multline}
are the dipole and octupole-induced intensity signals. If we rewrite the monopole-induced terms in equations \eqref{kSZIn1} and \eqref{kSZIn2} together, with the frequency and geometrical functions written out explicitly as

\begin{multline}\label{kSZIntot}
\frac{\Delta I_{\nu_c}^{\text{kSZ}} }{\Delta\tau} =
-\beta \cos\vartheta_\beta\Fnuc(T_0)\\
+\frac{1}{6}\beta^2(3\cos^2\vartheta_\beta -1)\Fnu(T_0)x\coth(x/2).
\end{multline}
we obtain the well known kSZ intensity (kSZIn) effect and its first relativistic correction in $\beta$. To first order in $\beta$, kSZIn is proportional to the parallel component of the peculiar velocity $\beta \cos\vartheta_\beta=\beta_\parallel$ as expected. Note that $\beta^3$ order corrections due to the leakage of the monopole into the $\delta I_{\nu_c}^{(3)}$ will be comparable to the motion-induced intensity effects  for clusters with large peculiar velocity and therefore must be included. The third and fourth lines of equations \eqref{T1inIn} and \eqref{T3inIn} are the dipole and octupole-induced intensity effects which we will collectively call T$_{1,3}$inIn. The top panel of figure \ref{fig:newtinIn} shows the total signal induced by T$_{2}$inIn and T$_{1,3}$inIn in the direction of the north galactic pole for an exaggerated value of $\beta=0.2$. It clearly shows the dependence of the frequency function of the total signal on the cluster's direction of the motion. The T$_{1,3}$inIn effect is maximal for clusters moving parallel to the line of sight. The bottom panel shows how these signals compare to each other.

The motion-induced intensity by the dipole and octupole change the angular distribution of the TinIn signal over different frequency bands. Figure \ref{fig:TIT-moving-217GHz} shows the absolute value of the TinIn signal for clusters moving with $\vec{\bm \beta}=-0.05 \hat{\bm{z}}$ at 217 GHz and how the angular shape of the signal changes at different frequencies. Similar to figure \ref{fig:TIP-moving-217GHz}, this plots only shows the angular distortions calculated for a particular direction of motion of the clusters and does not show the dependence of the angular distortions on $\betahat$. The change in the location of the maxima and minima at different frequencies are smaller compared to the TinPol effect but are still noticeable. As mentioned before, the $\eth^{\pm2}$ derivative of these maps will reproduce rescaled version of the polarization maps in figure \ref{fig:TIP-moving-217GHz} and are highly correlated with them. This correlation can be used to amplify the signal to noise ratio for both signals.

\section{\label{sec:IV}Summary and Discussion}

In this paper we performed a comprehensive study of the expected signal from CMB-gas interaction in moving galaxy clusters over the whole sky. We developed a formalism that generalizes the kSZ effect by including the CMB temperature anisotropies and their frequency dependences. In this formalism one can easily see the contribution of the low multipoles of the CMB to the intensity and polarization distortions induced in the direction of a galaxy cluster. In the absence of anisotropies our results naturally reduce to the well-known kSZ intensity and polarization effects. However, in the anisotropic picture the low multipoles of the CMB also contribute to the kSZ effect in a frequency dependent manner. For non-moving clusters the polarization signal is only proportional to the quadrupole of the temperature, but we showed that for clusters with large peculiar velocities the primordial dipole and octupole can contribute to the signal by 10\% (20\%) over 20\% (10\%) of the sky at 300 GHz. Measureing these intensity and polarization distortions in the direction of a cluster can be exploited to infer the low multipoles at other locations/redshifts in the universe and ultimately help to reduce the cosmic variance for these modes.  

In order to calculate the polarization induced by temperature anisotropies, we employed the aberration kernel formalism, which calculates the harmonic coefficients of anisotropies in a moving frame as a function of their CMB rest frame counterparts. We generalized this formalism in two ways: by allowing an arbitrary location and  direction of motion for the moving frame of the cluster, and by taking the frequency dependence of the anisotropies into account. Using these features, we developed a whole-sky and frequency-dependent formalism for calculating the kSZ effect. These generalizations reveal a connection between the frequency dependence of the signal and the direction of motion of the cluster for higher order effects. This will facilitate the interpretation of results from future microwave surveys, which are aiming to measure the frequency spectrum of the CMB over numerous channels, and allow for precise measurements of kSZ. Our generalizations of the aberration kernel can be also employed in the local frame to conveniently deboost/deaberrate the CMB multipoles in a frequency dependent manner. 

First we studied the temperature-induced polarization signal and its dependence on the remote quadrupole in the direction of non-moving clusters (\S \ref{sec:II}). By using our local temperature quadrupole, we presented a map of the expected quadrupole-induced polarization signal and its spatial morphology for non-moving clusters at $z\sim0$. The peak value of this signal for a single cluster is expected to be $\sim1.7\tau$ kJy/sr ($3.6 \tau \muK$) which results in a 70 nK signal for a cluster with $\tau=0.02$. The signal has a maximum at 218 GHz in the $(l,b)=(-113.1^{\circ},-63.3^{\circ})$ and $(67.6^{\circ},67.9^{\circ})$, which would be the ideal directions for its detection. At higher redshifts the amplitude and angular dependence of the expected signal will be different due to the change in the quadrupole, so the polarization map will be less correlated with the one calculated at $z\sim0$ and the local quadrupole \cite{Seto:2005de}. 

We then generalized the problem to the case of moving clusters and calculated the contribution of the quadrupole's neighbors to the signal (\S \ref{sec:III}), the leakage of the monopole into the quadrupole observed by the cluster leads to the kSZ polarization effect. This signal is subdominant to the quadrupole-induced polarization for typical values of cluster velocity ($\beta \lesssim 0.003$) but can become dominant for high velocity clusters and mergers. For example in the case of the bullet cluster ($\beta\approx0.015$) \citep{Markevitch:2003at} kSZ polarization is 25 times larger than the quadrupole-induced polarization at 218 GHz. Nevertheless, since the peak of kSZ polarization is always at 276 GHz, these effects are distinguishable from each other in multi-frequency surveys. One can also take advantage of the distinct patterns that these signals create in the $Q$ and $U$ polarization components, which are easily separable in our formalism. 

For a moving cluster the next order correction induced by the other anisotropies is due to the dipole and octupole moments of the CMB. Although the contribution of dipole and octupole to the induced quadrupole polarization are of order $\beta$, their individual frequency weights can enhance the signal by a factor of 5 and larger for $\nu\geq$ 350 GHz, so for $\beta=0.015$ they can collectively boost the polarization signal by 15\%.  For a smaller value of $\beta=0.003$ the change is about 2\%-10\% in the 200-600 GHz frequency range. Unlike kSZ and the quadrupole-induced polarization, the frequency function of the dipole and octupole-induced polarizations depend on the direction of motion of the cluster $\betahat$. This is due to the fact that the total signal is a combination of the Doppler (frequency-dependent) and aberration (frequency-independent) effects. Since the ratio of these effects depends on $\betahat$, different directions of motion change the peak location of the frequency function for the total induced signal. 

The quadrupole-induced polarization also has a different angular dependence compared to that of the dipole and octupole, so the ratio between the two signals changes over the sky. This will make the dipole and octupole-induced effects non-negligible in large areas over the sky. Moreover, since the contribution of the dipole and octupole to the overall polarization signal has a steeper frequency dependence than the one due to the quadrupole, they may become dominant at high frequencies ($\nu \gtrsim 400$ GHz). As for the next leading order term, the hexadecapole, since its leakage to the quadrupole is proportional to $\beta^2$, it only contributes to the signal by 0.2\% at low frequencies, but can be as large as 2\% at higher frequencies.

The temperature-induced polarization effects for the leading order terms can be summarized (in order of importance) as follows:  

\begin{itemize}[leftmargin=*]
	\item Monopole-induced polarization (kSZPol) for high velocity ($\beta\gtrsim0.003$) clusters and mergers $\propto \beta^2\tau a^T_{00}$
	\item Quadrupole-induced polarization (T$_2$inPol) $\propto \tau a^T_{2m}$
	\item Monopole-induced polarization (kSZPol) for low velocity ($\beta\lesssim0.003$) clusters $\propto \beta^2 \tau a^T_{00}$
	\item Dipole-induced polarization (T$_1$inPol) $\propto \beta \tau a^T_{1m}$
	\item Octupole-induced polarization (T$_3$inPol)  $\propto  \beta \tau a^T_{3m}$
	\item Hexadecapole-induced polarization	(T$_4$inPol) $\propto\beta^2\tau a^T_{4m}$
\end{itemize}   
It is important to mention that  for certain alignments of the line of sight and the velocity vector, the angular and geometrical prefactors ---which have not been included in the list--- can supress each term with respect to the others or even make it vanish. For example since the kSZPol effect only depends on the transverse component of $\beta$, it becomes zero for a cluster that has a bulk velocity parallel to our line of sight. Also regardless of the magnitude or direction of $\vec{\bm \beta}$, T$_2$inPol vanishes in four directions over the sky. So for clusters located close to these directions and moving with a radial bulk velocity, T$_1$inPol and T$_3$inPol are the dominant polarization effects. Note that since the late ISW effect enhances the dipole mode of the CMB more than the octupole mode, the induced polarization by the former is in general expected to be larger than that of the latter. This is why T$_1$inPol appears before T$_3$inPol in the above list.

The quadrupole of the CMB is also reflected through the temperature-induced intensity effect for a non-moving cluster with a maximum value of $\sim 9.1 \tau ~\text{kJy/sr}~ (18.7\tau ~\muK)$ which is about 5 times larger than the T$_2$inPol signal (\S \ref{sec:IIB}). Unfortunately, unlike the polarization effect, this signal does not directly probe the quadrupole moment of the CMB in the direction of non-moving clusters because it is confused by the other temperature multipoles that scatter out of line of sight. However, as discussed in \S \ref{sec:IIB2}, the second derivative of the intensity distortion map is highly correlated with the temperature-induced polarization so cross-correlating the two maps can amplify the signal to a great extent. 

The sensitivity needed for measuring the T$_2$inPol  ($\sim$ 100 nK) and T$_1$inPol and T$_3$inPol ($\sim$ 10 nK) are well below the sensitivity level of current instruments. The PRISM project \citep{Andre:2013nfa} has proposed to measure the quadrupole-induced polarization signal, and since it is a multi-frequency and whole-sky survey it is the ideal instrument to measure the signal induced by the other low multipoles as well. The sensitivity of PRISM at relevant frequency channels is not high enough for single cluster measurements, but the signal can be enhanced to detection level using stacking methods and cross-correlation with the temperature-induced intensity effects discussed earlier. 
 
The polarization induced by the low multipoles of the CMB, aside from imposing corrections to the kSZ effect, allow us to measure these modes at higher redshifts. Successful measurements of the low multipoles can help us determine if the observed anomalies in the quadrupole and octupole are coincidental or fundamental. Even more importantly, it opens up a window to find the primordial dipole moment of the CMB at $z\approx0$ which is inevitably masked by the motion-induced dipole in our local moving frame. In this study we neglected the multiple scattering events, which can induce similar effects in aspherical clusters ($\sim 0.2\%$) \citep{Chluba2013a}. We also neglected the initial polarization of the CMB and its leakage into the quadrupole polarization through galaxy clusters. This effect is generally expected to be small, however, in certain areas of the sky can be comparable to the temperature-induced polarization and kSZ. We will investigate the polarization-induced polarization effects in a future paper.  

\section{Acknowledgments}
SY cordially thanks Loris Colombo and Itzhak Bars for helpful discussions. The authors sincerely thank Jens Chluba for his detailed comments on the manuscript. We are also very grateful to the anonymous referee for their careful analysis of our work. We acknowledge using the Wolfram Mathematica 10.2 Software for preparing the plots. We also thank the WiSE program and the USC supercomputing center for their support. 
\bibliographystyle{apsrev4-1}	
\bibliography{kSZPol}

%\newpage
\appendix
\numberwithin{equation}{section}

\section{\label{sec:appA} Frequency Functions}
The assumption that the frequency spectrum of the CMB is isotropic was used in equation \eqref{alm(I)_to_alm(T)} to separate the frequency dependence of the multipoles from their spatial morphology. Assuming that the Intensity of the CMB can be described by a pure blackbody $B_\nu(T)\equiv\frac{2h\nu^3}{c^2}\frac{1}{e^{h\nu/kT}-1}$ in its rest frame, we can write 

\begin{align}\label{app:delta_I}
I_{\nu}(T)=&B_\nu(T)\\
\delta I_{\nu}(T)=&\frac{\partial B_\nu(T)}{\partial T}\delta T\\
=&T^{-1}F_{\nu}(T)\delta T
\end{align}
where $F_\nu(T)\equiv B_\nu(T) \frac{xe^{x}}{e^{x}-1}$ and $x=h\nu/kT$ is the dimensionless frequency. Here since $\delta T / T \approx 10^{-5}$ we can safely ignore the second and higher order temperature fluctuations. By using the expansions \eqref{intensity_expansion} and \eqref{temperature_expansion} on both sides of equation \eqref{app:delta_I} one can easily derive equation \eqref{alm(I)_to_alm(T)}, which shows that the frequency dependence of the CMB monopole is described by $B_\nu(T)$, and for the anisotropies by its first temperature derivative $F_\nu(T)$.
Since the aberration effect is frequency-independent and only depends on the angle between the incoming photon and observer's velocity vector, it does not change the frequency spectrum of the observed multipoles. Therefore the leakage of the monopole into the its nearby multipoles will be proportional to $B_\nu(T)$ and for the anisotropies to $F_\nu(T)$. The Doppler effect on the other hand shifts the observed frequency of the incoming photons in an angle dependent manner and is proportional to the derivatives of the frequency spectrum. Up to first order in $\beta$, the Doppler leakage of the multipoles is proportional to their first frequency derivative ($\Dnu^{(11)}$), and to second order in $\beta$ proportional to the second derivatives ($\Dnu^{(20)}$ and $\Dnu^{(22)}$) and so on. For the monopole the explicit form of these functions are 

\begin{align}
B^{(11)}_\nu(T)\equiv&\Dnu^{(11)}B_\nu(T)= -F_\nu(T),\\
B^{(20)}_\nu(T)\equiv&\Dnu^{(20)}B_\nu(T)= \frac{1}{6}F_\nu(T)(x\coth(x/2)-3),\\
B^{(22)}_\nu(T)\equiv&\Dnu^{(22)}B_\nu(T)= \frac{1}{3}F_\nu(T)x\coth(x/2),
\end{align}
and for the anisotropies 

\begin{align}
F^{(11)}_\nu(T)\equiv&\Dnu^{(11)}F_\nu(T)= -F_\nu(T)(x\coth(x/2)-1),\\
F^{(20)}_\nu(T)\equiv&\Dnu^{(20)}F_\nu(T)= \frac{1}{6}F_\nu(T)\\
\times&\frac{-3+2x^2+(3+x^2)\cosh(x)-5x\sinh(x)}{\cosh(x)-1}, \nonumber\\
F^{(22)}_\nu(T)\equiv&\Dnu^{(22)}F_\nu(T)= \frac{1}{6}F_\nu(T)x\sinh^{-2}(x/2)\\ &\hspace{2em}\times (x\cosh(x)-2\sinh(x)+2x). \nonumber
\end{align}

\begin{figure}
	\centering
	\includegraphics[width=1.05\linewidth]{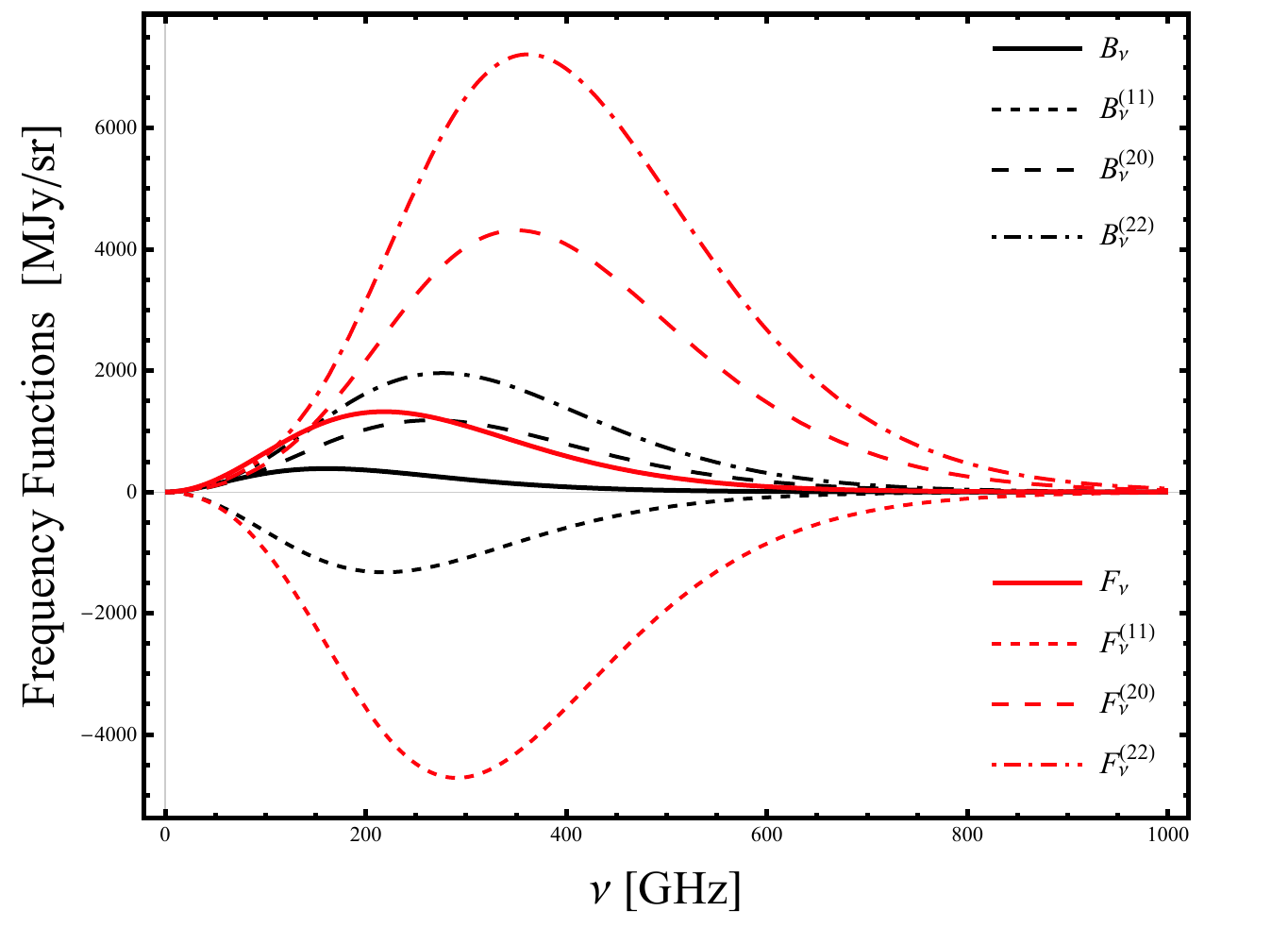}
	\caption{Frequency functions of the CMB monopole ($B_\nu$), its anisotropies ($F_{\nu}$) and their derivatives. The large amplitude of the derivatives boosts the Doppler leakage of the multipoles into each other.}
	\label{fig:FPlots}
\end{figure}

Figure \ref{fig:FPlots} shows all these frequency functions together. Note that the frequency functions used in \S \ref{sec:II} and \S \ref{sec:III} are normalized with the temperature monopole at redshift $z$ and are denoted with a tilde. Notice that the frequency dependence of the primordial dipole $F_{\nu}$ is identical to the frequency dependence of the Doppler leakage of the monopole $B_\nu^{(11)}$ up to a negative sign. This is why the primordial dipole moment of the temperature anisotropies is not distinguishable from the dipole induced by the monopole, due to our motion with respect to the CMB rest frame. 

\section{Calculation Details}
\subsection{\label{sec:appB1}The Intensity Kernel Integral}
The harmonic coefficients in the cluster's frame can be found by performing the following integral from equation \eqref{alm_cluster} 

 \begin{multline}\label{app:alm_integral}
 a^{I_c}_{\ell' m'}(\nu_c) =\\
 \sum_{\ell=0}^{\infty}\sum\nolimits_{m}^{\ell} \int \Big(\frac{\nu_c}{\nu_{cmb}}\Big)^3 a^{I}_{\ell m}(\nu_{cmb}) Y_{\ell m}(\gammahat_{cmb}) Y^*_{\ell' m'}(\gammahat_c) d^2 \gammahat_c.
 \end{multline}
 For small values of $\beta$ we can use the inverse of equations \eqref{doppler_freq} and \eqref{aberration} and perform the following Taylor expansions 

\begin{multline}\label{app:alm_taylor_expand}
a^{I}_{\ell m}(\nu_{cmb}) = a^{I}_{\ell m}(\nu_c) +\\
\left(\mu_c \beta+\frac{1}{2}\beta^2 \right) 
\nu_c \partial_{\nu}\Big|_{\nu=\nu_c} a^{I}_{\ell m}(\nu)+\\
\left(\frac{1}{2}\mu_c^2 \beta^2 \right) \nu_c^{2}\partial^2_{\nu}\Big|_{\nu=\nu_c} a^{I}_{\ell m}(\nu)+O(\beta^3),
\end{multline}

\begin{multline}\label{app:Ylm_taylor_expand}
Y_{\ell m}(\gammahat_{cmb})= Y_{\ell m}(\gammahat_c) +\\ \left(\beta+\frac{1}{2}\mu_c\beta^2\right)\betahat.\nabla Y_{\ell m}(\gammahat_c)+\\ \frac{1}{2}\beta^2 \left(\betahat.\nabla\right)^2 Y_{\ell m}(\gammahat_c)+O(\beta^3).
\end{multline}
where $\mu_c=\gammahat_c.\betahat$. The expansions are respectively due to the Doppler and aberration effects. In the following subsections we simplify the RHS of these equations analytically and integrate equation \eqref{app:alm_integral}.

\subsubsection{\label{sec:appB1a}Doppler effect}
It is mathematically convenient to write different powers of $\mu$ in equations \eqref{app:alm_taylor_expand} and \eqref{app:Ylm_taylor_expand} in terms of the Legendre polynomials as $\mu_c=P_1(\gammahat_c.\betahat)$ and $\mu_c^2=\frac{1}{3}+\frac{2}{3}P_2(\gammahat_c.\betahat)$. This allows us to use the addition theorem for spherical harmonics, as a special case of equation \eqref{addition}

\begin{equation}
P_{\ell}(\gammahat_c.\betahat)=\frac{4\pi}{2\ell+1}\sum\nolimits_{n}^{\ell} Y_{\ell n}(\gammahat_c) Y^*_{\ell n}(\betahat),
\end{equation}
and separate the $\gammahat_c$ and $\betahat$ dependence and easily integrate over $\gammahat_c$ in equation \eqref{app:alm_integral}. We rewrite the expansion \eqref{app:alm_taylor_expand} as

\begin{multline}\label{app:alm_integrand_expand}
\left(\frac{\nu_c}{\nu_{cmb}}\right)^3a^{I}_{\ell m}(\nu_{cmb})=
\Dnu^{(00)}a^I_{\ell m}(\nu) \\
+\beta \Dnu^{(11)}a^I_{\ell m}(\nu)\sum\nolimits_{n}^{1}\frac{4\pi}{3}Y_{1n}(\gammahat_c)Y^*_{1n}(\betahat)\\
+\beta^2\Dnu^{(22)}a^I_{\ell m}(\nu)\sum\nolimits_{n}^{2}\frac{4\pi}{5}Y_{2n}(\gammahat_c)Y^*_{2n}(\betahat)\\
+\beta^2\Dnu^{(20)}a^I_{\ell m}(\nu)+O(\beta^3),
\end{multline} 
with the differential operators $\Dnu^{(kj)}$ defined as 

\begin{subequations}\label{app:derivatives}
	\begin{align}
	\Dnu^{(00)}&\equiv \partial_{\nu_c}^0,\\
	\Dnu^{(11)}&\equiv -3\partial_{\nu_c}^0+\nu_c\partial_{\nu_c}^1,\\
	\Dnu^{(20)}&\equiv \frac{1}{2}\partial_{\nu_c}^0-\frac{1}{2}\nu_c\partial_{\nu_c}^1+\frac{1}{6}\nu_c^2\partial_{\nu_c}^2,\\
	\Dnu^{(22)}&\equiv 4\partial_{\nu_c}^0-2\nu_c\partial_{\nu_c}^1+\frac{1}{3}\nu_c^2\partial_{\nu_c}^2,
	\end{align} 	
\end{subequations}
where $\partial_{\nu_c}^k\equiv \partial^k / \partial\nu^k |_{ \nu = \nu_c}$. The trivial operator $\Dnu^{(00)}$ only changes the argument of its following function from $\nu$ to $\nu_c$ and it is defined for consistency in notation.

\subsubsection{\label{sec:appB1b}Aberration effect}
The gradient terms in equation \eqref{app:Ylm_taylor_expand} can be simplified in a coordinate system where $\betahat=\hat{\bm z}$ using the recursive spherical harmonics identities
\begin{align}
(\mu^2-1)\frac{\partial}{\partial \mu}Y_{\ell m}=&\ell C_{(\ell+1) m}Y_{{\ell+1} m}-(\ell+1)C_{\ell m}Y_{(\ell-1)m},\\
\mu Y_{\ell m}=&C_{(\ell+1) m}Y_{(\ell+1) m}+C_{\ell m}Y_{(\ell-1)m}.
\end{align}
with the constants $C_{\ell m}$ defined as
\vspace{-0.5em}
\begin{equation}
C_{\ell m}\equiv\sqrt{\frac{\ell^2-m^2}{4\ell^2-1}}.
\end{equation}
In order to use these identities, first we rotate the cluster's coordinate system $(\hat{\bm{x}},\hat{\bm{y}},\hat{\bm{z}})$ to align $\hat{\bm z}$ with $\betahat$, then simplify the equations, and finally rotate it back to its original orientation. By applying an active rotation using the Euler matrix $R\equiv R_{\hat{\bm z}}(-\phi_\beta)R_{\hat{\bm y}}(-\theta_\beta)$ to the argument of equation \eqref{app:Ylm_taylor_expand}, we can rewrite it as
\begin{multline}\label{app:Ylm_aberration_simple}
Y_{\ell m}(R\gammahat_{cmb})= Y_{\ell m}(R\gammahat_c)\\ +\beta\mathcal{Y}^{(1)}_{\ell m}(R\gammahat_c)+\frac{1}{2}\beta^2 \mathcal{Y}^{(2)}_{\ell m}(R\gammahat_c)+O(\beta^3),
\end{multline}
where the functions $\mathcal{Y}^{(i)}_{\ell m}$ comprise a linear combination of $Y_{\ell m}$ and its $i^{\text{th}}$ nearest harmonic neighbors 
\begin{equation}
\mathcal{Y}^{(i)}_{\ell m}(\gammahat)\equiv\sum\nolimits_{j}^{i}\CEu^{(i,j)}_{\ell m}Y_{(\ell+j)m}(\gammahat).
\end{equation}
The only nonvanishing coefficients $\CEu^{(i,j)}_{\ell m}$ in equation \eqref{app:Ylm_aberration_simple} are  
\begin{subequations}
	\begin{align}
	\CEu^{(1,+1)}_{\ell m}=&~\ell C_{(\ell+1)m},\\
	\CEu^{(1,-1)}_{\ell m}=&-(\ell+1)C_{\ell m},\\
	\CEu^{(2,+2)}_{\ell m}=&~\ell(\ell+1)C_{(\ell+1)m}C_{(\ell+2)m},\\
	\CEu^{(2,0)}_{\ell m}~~=&~-(\ell^2-1)C^2_{\ell m}-\ell(\ell+2)C^2_{(\ell+1)m},\\
	\CEu^{(2,-2)}_{\ell m}=&~\ell(\ell+1)C_{(\ell-1)m}C_{\ell m},
	\end{align}
\end{subequations}
and the coefficients $\CEu^{(1,0)}_{\ell m}$, $\CEu^{(2,+1)}_{\ell m}$, and $\CEu^{(2,-1)}_{\ell m}$ are equal to zero. In order to write equation \eqref{app:Ylm_aberration_simple} in terms of $Y_{\ell m}(\gammahat)$ first we use the Wigner D matrix to separate the rotation operator 
\begin{equation}
Y_{\ell M}(R\gammahat)= \sum_{M'}D^\ell_{M'M}(-\phi_\beta,-\theta_\beta,0)Y_{\ell M'}(\gammahat),
\end{equation}
which followed by a further transformation 
\begin{equation}
\sum_{M} D^\ell_{Mm}(0,\theta_\beta,\phi_\beta)Y_{\ell M}(R\gammahat)=Y_{\ell m}(\gammahat)
\end{equation}
will give us the desired equation for aberration effect in terms of $Y_{\ell m}(\gammahat)$ 

\begin{multline}\label{app:Ylm_integrand_expand}
Y_{\ell m}(\gammahat_{cmb})=Y_{\ell m}(\gammahat_c)\\
+\beta\sum\nolimits_{j,M,M'}^{1,\ell,\ell+j}  D^{\ell}_{Mm}(\betahat)\CEu^{(1,+j)}_{\ell M}D^{\ell+j}_{M'M}(\betahat^{-1})Y_{(\ell+j)M'}(\gammahat_c)  \\
+\frac{\beta^2}{2} \sum\nolimits_{j,M,M'}^{2,\ell,\ell+j} D^{\ell}_{Mm}(\betahat)\CEu^{(2,+j)}_{\ell M}D^{\ell+j}_{M'M}(\betahat^{-1})Y_{(\ell+j)M'}(\gammahat_c)\\+
O(\beta^3).
\end{multline}
here $\betahat$ and $\betahat^{-1}$ in the argument of the Wigner D symbols are shorthand notation for the Euler angles $(0,\theta_\beta,\phi_\beta)$ and $(-\phi_\beta,-\theta_\beta,0)$. For negative values of $j$ the Wigner D matrix is not defined for $l+j<|M|<l$, however in favor of consistency of notation we resolve this issue by setting $D^{\ell+j}_{M'M}$ to zero for these values of $M$. 

\subsubsection{\label{sec:appB1c} The Kernel Integral}
Substituting equations \eqref{app:alm_integrand_expand} and \eqref{app:Ylm_integrand_expand} into \eqref{app:alm_integral} will yield

 \begin{align}
 a^{I_c}_{\ell' m'}(\nu_c) =& \mathcal{I}^0 \nonumber\\
 + & \beta (\mathcal{I}^{D1} +\mathcal{I}_{A1} ) \nonumber\\
 + &\beta^2 (\mathcal{I}^{D0}+ \mathcal{I}^{D2} + \mathcal{I}_{A2} +  \mathcal{I}^{D1}_{A1}  )\nonumber\\
 +&O(\beta^3)
 \end{align}
 where 
 \begin{multline}\label{zeroth_integral}
  \mathcal{I}^0 \equiv \sum_{\ell=0}^{\infty}\sum\nolimits_{m}^{\ell}\Dnu^{(00)}  a^{I}_{\ell m}(\nu)\\
  \times \int Y_{\ell m}(\gammahat_c) Y^*_{\ell' m'}(\gammahat_c)\text{d}^2 \gammahat_c
  \end{multline}
 is the zeroth order integral, 
  \begin{multline}
  \mathcal{I}^{D1} \equiv \sum_{\ell=0}^{\infty}\sum\nolimits_{m,n}^{\ell,1}\frac{4\pi}{3} Y^*_{1n}(\betahat)\Dnu^{(11)}  a^{I}_{\ell m}(\nu)  \\
  \times \int Y_{\ell m}(\gammahat_c) Y^*_{\ell' m'}(\gammahat_c)Y_{1 n}(\gammahat_c)  \text{d}^2 \gammahat_c
  \end{multline}
  is the first order Doppler integral,
   \begin{multline}
   \mathcal{I}_{A1} \equiv \sum_{\ell=0}^{\infty}\sum\nolimits_{m,j,M'}^{\ell,1,\ell+j}\Omega^{(1,j)}_{\ell m M'}(\betahat) \Dnu^{(00)}  a^{I}_{\ell m}(\nu) \\
   \times \int Y_{(\ell+j) M'}(\gammahat_c) Y^*_{\ell' m'}(\gammahat_c) \text{d}^2 \gammahat_c
   \end{multline}
   is the first order aberration integral,
    \begin{multline}
    \mathcal{I}^{D0} \equiv \sum_{\ell=0}^{\infty}\sum\nolimits_{m}^{\ell}\Dnu^{(20)}  a^{I}_{\ell m}(\nu) \\
    \times \int Y_{\ell m}(\gammahat_c) Y^*_{\ell' m'}(\gammahat_c)  \text{d}^2 \gammahat_c
    \end{multline}
    and 
     \begin{multline}
     \mathcal{I}^{D2} \equiv \sum_{\ell=0}^{\infty}\sum\nolimits_{m,n}^{\ell,2}\frac{4\pi}{5} Y^*_{2n}(\betahat)\Dnu^{(22)}  a^{I}_{\ell m}(\nu) \\
     \times \int Y_{\ell m}(\gammahat_c) Y^*_{\ell' m'}(\gammahat_c)Y_{2 n}(\gammahat_c)  \text{d}^2 \gammahat_c
     \end{multline}
     are the second order Doppler integrals,
      \begin{multline}
      \mathcal{I}_{A2} \equiv \sum_{\ell=0}^{\infty}\sum\nolimits_{m,j,M'}^{\ell,2,\ell+j}\frac{1}{2} \Omega^{(2,j)}_{\ell m M'}(\betahat) \Dnu^{(00)}  a^{I}_{\ell m}(\nu) \\
      \times \int Y_{(\ell+j) M'}(\gammahat_c) Y^*_{\ell' m'}(\gammahat_c)  \text{d}^2 \gammahat_c
      \end{multline}
      is the second order aberration integral and
 \begin{multline}\label{D1A1_Integral}
 	\mathcal{I}^{D1}_{A1} \equiv \sum_{\ell=0}^{\infty}\sum\nolimits_{m,n,j,M'}^{\ell,1,1,\ell+j}Y^*_{1n}(\betahat)\Omega^{(1,j)}_{\ell m M'}(\betahat) \Dnu^{(11)}  a^{I}_{\ell m}(\nu) \\
 	\times \frac{4\pi}{3} \int Y_{(\ell+j) M'}(\gammahat_c) Y^*_{\ell' m'}(\gammahat_c)Y_{1 n}(\gammahat_c)  \text{d}^2 \gammahat_c
 \end{multline}
is the cross-Doppler-aberration integral of first order. The rotation coefficients $\Omega^{(1,j)}_{\ell m M'}(\betahat)$ which were introduced to simplify the aberration expansion are defined as 

\begin{equation}\label{Omega_def}
\Omega^{(k,j)}_{\ell m M'}(\betahat)\equiv \sum\nolimits_{M}^{\ell} D^{\ell}_{Mm}(\betahat)\CEu^{(k,j)}_{\ell M}D^{(\ell+j)}_{M'M}(\betahat^{-1}).
\end{equation}
These integrals are easy to evaluate using
  \begin{equation}
  \int Y_{\ell m}(\gammahat_c) Y^*_{\ell' m'}(\gammahat_c)\text{d}^2 \gammahat_c = \delta_{\ell \ell'} \delta_{m m'}
  \end{equation}
and the Gaunt formula
 \begin{multline}\label{Gaunt}
 \int Y_{\ell m}(\gammahat_c) Y^*_{\ell' m'}(\gammahat_c)Y_{N n}(\gammahat_c)  \text{d}^2 \gammahat_c = \\
 =(-1)^{m'}\sqrt{\frac{(2\ell+1)(2\ell'+1)(2N+1)}{4\pi}} \\
 \times \begin{pmatrix}
 \ell& \ell'& N\\
 0& 0& 0\\
 \end{pmatrix}
 \begin{pmatrix}
 \ell& \ell'& N\\
 m& -m'& n\\
 \end{pmatrix}\\
 \equiv \Delta_{N,n}(\ell,m;\ell',m').
 \end{multline}
After substituting these into equations \eqref{zeroth_integral}-\eqref{D1A1_Integral} we can simplify them as 

 \begin{align}
 \label{app:I0}\mathcal{I}^0 =& \Dnu^{(00)}  a^{I}_{\ell' m'}(\nu)\\
 \mathcal{I}^{D0} =& \Dnu^{(20)}  a^{I}_{\ell' m'}(\nu)\\
 \mathcal{I}^{Dp}_{Aq} =& \label{app:IDA}\sum_{\ell=0}^{\infty}\sum\nolimits_{m}^{\ell}\Gij{p}{q}^{\ell,m}_{\ell 'm'}(\betahat)\Dnu^{(pp)}a^{I}_{\ell m}(\nu)
 \end{align}
where the geometrical factors $\Gij{p}{q}^{\ell,m}_{\ell 'm'}(\betahat)$ are defined as 

\begin{align}\label{Gij}
 \Gij{1}{0}^{\ell,m}_{\ell 'm'}(\betahat)=& \sum\nolimits_{n}^{1}\frac{4\pi}{3} Y^*_{1n}(\betahat) \Delta_{1,n}(\ell,m;\ell',m')\\
\Gij{0}{1}^{\ell,m}_{\ell 'm'}(\betahat)=&\sum\nolimits_{j,M'}^{1,\ell+j}\Omega^{(1,j)}_{\ell m M'}(\betahat) 
\delta_{\ell+j,\ell'}\delta_{M',m'}\\
\Gij{2}{0}^{\ell,m}_{\ell 'm'}(\betahat)=& \sum\nolimits_{n}^{2}\frac{4\pi}{5} Y^*_{2n}(\betahat) \Delta_{2,n}(\ell,m;\ell',m')\\
\Gij{0}{2}^{\ell,m}_{\ell 'm'}(\betahat)=&\sum\nolimits_{j,M'}^{2,\ell+j}\Omega^{(2,j)}_{\ell m M'}(\betahat) \delta_{\ell+j,\ell'}\delta_{M',m'}\\
 \Gij{1}{1}^{\ell,m}_{\ell 'm'}(\betahat)=&\sum\nolimits_{n,j,M'}^{1,1,\ell+j}\frac{4\pi}{3}Y^*_{1n}(\betahat)\Omega^{(1,j)}_{\ell m M'}(\betahat) \label{Gij2}\\ &\hspace{4em}\times\Delta_{1,n}(\ell+j,M';\ell',m') \nonumber
\end{align}

Here in the Doppler terms we have introduced the notation $\Delta_{N,n}(\ell,m;\ell',m')$
which can be easily calculated in terms of the Wigner 3-j symbols using equation \eqref{Gaunt}. These coefficients are nonzero for $|\ell-\ell'|<N<\ell+\ell'$, so up to order $\beta^N$, $\Delta_{N,n}(l,m;l',m')$ brings in contributions from the first $N$ neighbors of $\ell^{\text{th}}$ multipole to the observed multipoles of order $\ell'$ with different frequency weights. Similarly the order $\beta^N$ aberration effect draws in the first $N$ neighbors of each multipole, but unlike the Doppler effect, with the same frequency weight as the observed multipole. 

\begin{widetext}
%\pagebreak
\subsection{\label{sec:appBII}The Observed Multipoles in a Boosted Frame}
After substituting equations \eqref{app:I0}-\eqref{app:IDA} in \eqref{app:alm_integral}, we obtain the following expression for the  $a^{I_c}_{lm}$s

	\begin{equation}\label{final_alms2}
	\begin{split}
	a^{I_c}_{\ell'm'}(\nu_c)=& a^{I}_{\ell'm'}(\nu_c)+\\
	\text{Doppler}	
	&\begin{cases}
	+\beta~\sum\limits_{\ell=0}^{\infty}\sum\nolimits_{m,n}^{\ell,1}~
	\frac{4\pi}{3}Y^*_{1n}(\hat{\bm{\beta}})\Delta_{1,n}(\ell,m;\ell',m')\Dnu^{(11)}a^{I}_{\ell m}(\nu)\\
	+\beta^2~\Dnu^{(20)}a^{I}_{\ell'm'}(\nu)\\
	+\beta^2\sum\limits_{\ell=0}^{\infty}\sum\nolimits_{m,n}^{\ell,2}\frac{4\pi}{5}
	Y^*_{2n}(\hat{\bm{\beta}})\Delta_{2,n}(\ell,m;\ell',m')\Dnu^{(22)}a^{I}_{\ell m}(\nu)
	\end{cases}\\
	\text{Aberration} 
	&\begin{cases}
	+\beta \sum\limits_{\ell=0}^{\infty}\sum\nolimits_{m,j,M',M}^{\ell,1,\ell+j,\ell} D^{\ell}_{Mm}(\betahat)\CEu^{(1,j)}_{\ell M}D^{(\ell+j)}_{M'M}(\betahat^{-1})
	\delta_{\ell+j,\ell'}\delta_{M',m'}\Dnu^{(00)}a^{I}_{\ell m}(\nu)\\
	+\frac{1}{2}\beta^2 \sum\limits_{\ell=0}^{\infty}\sum\nolimits_{m,j,M',M}^{\ell,2,\ell+j,\ell} D^{\ell}_{Mm}(\betahat)\CEu^{(2,j)}_{\ell M}D^{(\ell+j)}_{M'M}(\betahat^{-1})
	\delta_{\ell+j,\ell'}\delta_{M',m'}\Dnu^{(00)}a^{I}_{\ell m}(\nu)\\
	\end{cases}\\
	\text{Doppler+Aberration}
	&\begin{cases}
	+\beta^2 \sum\limits_{\ell=0}^{\infty}\sum\nolimits_{m,n,j,M',M}^{\ell,1,1,\ell+j,\ell} \frac{4\pi}{3}Y^*_{1n}(\betahat)D^{\ell}_{Mm}(\betahat)\CEu^{(1,j)}_{\ell M}D^{(\ell+j)}_{M'M}(\betahat^{-1})
	\Delta_{1,m''}(\ell+j,M';\ell',m')\\
	\qquad \times \Dnu^{(11)}a^{I}_{\ell m}(\nu)
	+O(\beta^3).\\
	\end{cases}
	\end{split}
	\end{equation}

Using this equation we can easily find the low multipoles of the CMB in any boosted frame. The source of each term is written on the left for clarification. It is easy to see from this expression that the Doppler leakage changes the frequency function of the multipoles in the moving frame. Equation \eqref{final_alms2} can be used in our local frame to disentangle the leakage of the multipoles into each other due to our motion with respect to the CMB rest frame. Since we derived this equation for an arbitrary direction of motion, it gives us the advantage to treat $\betahat$ as an independent parameter when we are deboosting the observed multipoles. In the following subsections we calculate the observed monopole, dipole and quadrupole in a frame moving with the velocity vector $\vec{\bm{\beta}}=\beta \betahat$. 

\subsubsection{Monopole \emph{(}$\ell'=0$\emph{)}}
The expression for the observed monopole in the cluster's frame to second order in $\beta$ can be simplified as 
	\begin{equation}\label{app:final_monopole}
	\begin{split}
	a^{I_c}_{00}(\nu_c)=& \Bnuc(T_0)a^{T}_{00}+\\
	\text{Doppler}	
	&\begin{cases}
	+\beta^2\Bnuc^{(20)}(T_0)a^T_{00}\\
	+\beta~\sum\nolimits_{m}^{1}
	\frac{2\sqrt{\pi}}{3}Y_{1m}(\hat{\bm{\beta}})\Fnuc^{(11)}(T_0)a^{T}_{1 m}\\
	+\beta^2\sum\nolimits_{m}^{2}
	\frac{2\sqrt{\pi}}{5}Y_{2m}(\hat{\bm{\beta}})\Fnuc^{(22)}(T_0)a^{T}_{2 m}\\
	\end{cases}\\
	\text{Aberration} 
	&\begin{cases}
	-\beta~ \sum\nolimits_{m}^{1} \frac{4\sqrt{ \pi}}{3}Y_{1m}(\betahat)\Fnuc(T_0)a^{T}_{1 m}\\
	+\beta^2 \sum\nolimits_{m}^{2} \frac{4\sqrt{ \pi}}{5}Y_{2m}(\betahat)\Fnuc(T_0) a^{T}_{2 m}
		+O(\beta^3).
	\end{cases}
	\end{split}
	\end{equation}
It is evident from this expression that the frequency spectrum of the new monopole is different from the spectrum of the monopole in the CMB frame $a^I _{00}(\nu)$ which can be described with $\Bnuc(T_0)$. We calculated the aberration kernel, based on the assumption that the intensity of the CMB in every direction of the sky can be described by a pure blackbody at temperature $T_0$. But in a moving frame, because of the Doppler effect, the temperature of these blackbodies will be different in different directions. Therefore, at each frequency band the average of intensity over the whole sky (monopole) picks up different values from different directions. The sum of these values do not necessarily reproduce a blackbody spectrum \citep{Chluba:2012gq,Chluba:2004cn,Khatri:2012rt}. For example in the absence of the anisotropies, a moving observer measures an intensity Doppler distortion of order $\delta I \approx \beta \partial_{\nu} \Bnu a^T_{00}$ in the forward direction and $-\delta I$ in the opposite direction. Averaging the intensity over all incoming angles will cancel the two distortions, so the observed monopole will be still a blackbody to first order in $\beta$. However, to second order in $\beta$ there will be Doppler distortions of order $\beta^2 \partial^2_{\nu} \Bnu a^T_{00}$ in all directions perpendicular to the direction of motion, that do not cancel each other in the average. Therefore, the new monopole observed in the moving frame will have a different frequency spectrum than the monopole of a pure blackbody radiator \citep{Chluba:2016bvg}. 

\subsubsection{Dipole \emph{(}$\ell'=1$\emph{)}}
The new dipole in the moving frame is
	\begin{equation}\label{app:final_dipole}
	\begin{split}
	a^{I_c}_{1m'}(\nu_c)=& \Fnuc(T_0)a^{T}_{1m'}+\\
	\text{Doppler}	
	&\begin{cases}
	-\beta ~
	\frac{2\sqrt{\pi}}{3}Y^*_{1m'}(\hat{\bm{\beta}})\Fnuc(T_0)a^{T}_{00}\vspace{0.5em}\\
	+\beta \sum\nolimits_{m,n}^{2,1}
	\sqrt{\frac{8\pi}{3}}Y^*_{1n}(\hat{\bm{\beta}})(-1)^{m'}\threeJ{2}{1}{1}{m}{-m'}{n}\Fnuc^{(11)}(T_0)a^{T}_{2m}\vspace{0.5em}\\
	+\beta^2~\Fnuc^{(20)}(T_0)a^{T}_{1m'}\vspace{0.5em}\\
	+\beta^2 \sum\nolimits_{m,n}^{1,2}
	\frac{2\sqrt{6\pi}}{5}Y^*_{2n}(\hat{\bm{\beta}})(-1)^{m'}\threeJ{1}{1}{2}{m}{-m'}{n}\Fnuc^{(22)}(T_0)a^{T}_{1m}\vspace{0.5em}\\
	-\beta^2 \sum\nolimits_{m,n}^{3,2}
	\frac{6\sqrt{\pi}}{5}Y^*_{2n}(\hat{\bm{\beta}})(-1)^{m'}\threeJ{3}{1}{2}{m}{-m'}{n}\Fnuc^{(22)}(T_0)a^{T}_{3m}\vspace{0.5em}\\
	\end{cases}\\
	\text{Aberration} 
	&\begin{cases}
	+\beta ~\sum\nolimits_{m,M}^{2,1} D^{2}_{Mm}(\betahat)\CEu^{(1,-1)}_{2 M}D^{1}_{m'M}(\betahat^{-1})
	\Fnuc(T_0)a^{T}_{2 m}\\
	+\frac{1}{2}\beta^2 \sum\nolimits_{m,M}^{1,1} D^{1}_{Mm}(\betahat)\CEu^{(2,0)}_{1 M}D^{1}_{m'M}(\betahat^{-1})
	\Fnuc(T_0)a^{T}_{1 m}\\
	+\frac{1}{2}\beta^2 \sum\nolimits_{m,M}^{3,1} D^{3}_{Mm}(\betahat)\CEu^{(2,-2)}_{3 M}D^{1}_{m'M}(\betahat^{-1})
	\Fnuc (T_0)a^{T}_{3 m}	
	\end{cases}\\
	\text{Doppler+Aberration}
	&\begin{cases}
	+     \beta^2 \sum\nolimits_{m,n,M',M}^{1,1,2,1} \sqrt{\frac{8\pi}{3}}Y^*_{1n}(\betahat)D^{1}_{Mm}(\betahat)\CEu^{(1,1)}_{1 M}D^{2}_{M'M}(\betahat^{-1}) (-1)^{m'}\threeJ{2}{1}{1}{M'}{-m'}{n}\Fnuc^{(11)}(T_0)a^T_{1m}
	\vspace{0.5em}\\
	+     \beta^2 \sum\nolimits_{m,n,M',M}^{3,~1,~2,~2} \sqrt{\frac{8\pi}{3}}Y^*_{1n}(\betahat)D^{3}_{Mm}(\betahat)\CEu^{(1,-1)}_{3 M}D^{2}_{M'M}(\betahat^{-1}) (-1)^{m'}\threeJ{2}{1}{1}{M'}{-m'}{n}\Fnuc^{(11)}(T_0)a^T_{3m}
	\vspace{0.5em}\\ 
	+     \beta^2 \sum\nolimits_{m,n}^{1,1} \frac{4\pi}{3}Y^*_{1n}(\betahat)Y^*_{1m}(\betahat)(-1)^{m'}\threeJ{0}{1}{1}{0}{-m'}{n}\Fnuc^{(11)}(T_0)a^T_{1m}+O(\beta^3).\\
	\end{cases}
	\end{split}
	\end{equation}
	Since the CMB monopole is the dipole's first neighbor, its Doppler leakage is proportional to $\beta$. Since the monopole is larger than the anisotropies by a factor of $10^5$, it is the dominant term in the new dipole, even for small values of $\beta$. Most importantly, the frequency weight of this term is exactly the same as the primordial dipole. Therefore, by only looking at the dipole moment of the CMB in a moving frame, one cannot distinguish between the  primordial dipole and the one induced by the monopole. Using this expression in our local coordinate system makes it clear that a bulk velocity of about $v\approx300$ km/s will induce a spurious dipole of order $10^{-3}\times 3 \text{K}=3\text{mK}$. This is precisely why the observed dipole in our coordinate system is always associated with our bulk motion in the CMB frame. However, since the Doppler leakage of the primordial dipole moment into its first neighbors has a different frequency weight, it can be extracted from them using the frequency spectral distortions of the observed monopole and quadrupole. It is important to mention that since the CMB monopole is isotropic by definition, the aberration leakage of this term into the other multipoles always vanishes as expected.

\subsubsection{\label{sec:appBIIc}Quadrupole \emph{(}$\ell'=2$\emph{)}}
 Similar to the monopole and dipole, we can easily calculate the observed quadrupole moment in the cluster's moving frame. Since the polarization induced in the direction of a cluster reflects the quadrupole that it observes, all the multipoles that leak into this mode will be observable in the polarization signal. The leakages of the first neighbors of the quadrupole, the dipole and octupole, are proportional to $\beta$ but are amplified by the frequency weight $\Fnuc^{(11)}(T_0)$ in the Doppler terms. The contribution of the second neighbors, the monopole and hexadecapole, is of order $\beta^2$. Although the frequency weight of the hexadecapole is much larger than the dipole and octupole, its dependence on the velocity keeps it sub-dominant for typical values of the cluster peculiar velocity. The monopole-induced quadrupole term on the other hand is much larger due to its dependence on $a^T_{00}$, so its leakage can become dominant over the other modes. This term is typically referred to as the \emph{Doppler Quadrupole} \citep{Kamionkowski:2002nd,Notari:2015kla,Chluba:2004cn}; In the context of galaxy clusters this term induces the kSZ polarization and the first relativistic correction of the kSZ intensity effect.  
		\begin{equation}\label{final_alms}
		\begin{split}
		a^{I_c}_{2 m'}(\nu_c)=& \Fnuc(T_0)a^{T}_{2 m'}+\\
		\text{Doppler}	
		&\begin{cases}
		+\beta^2 ~\frac{2\sqrt{\pi}}{5}Y^*_{2m'}(\betahat)\Bnuc^{(22)}(T_0)a^T_{00}\\
		+\beta \sum\nolimits_{m,n}^{1,1}~
		\sqrt{\frac{8\pi}{3}}Y^*_{1n}(\hat{\bm{\beta}})(-1)^{m'}\threeJ{1}{2}{1}{m}{-m'}{n}\Fnu^{(11)}(T_0)a^{T}_{1 m}\\
		-\beta \sum\nolimits_{m,n}^{3,1}~
		\sqrt{2 \pi} Y^*_{1n}(\hat{\bm{\beta}})(-1)^{m'}\threeJ{3}{2}{1}{m}{-m'}{n}\Fnu^{(11)}(T_0)a^{T}_{3 m}\\
		+\beta^2~\Fnuc^{(20)}(T_0)a^{T}_{2 m'}\\
		-\beta^2 \sum\nolimits_{m,n}^{2,2}~
		\sqrt{\frac{8\pi}{7}}Y^*_{2n}(\hat{\bm{\beta}})(-1)^{m'}\threeJ{2}{2}{2}{m}{-m'}{n}\Fnu^{(22)}(T_0)a^{T}_{2 m}\\
		+\beta^2 \sum\nolimits_{m,n}^{4,2}~
		\frac{6\sqrt{2\pi}}{\sqrt{35}}Y^*_{2n}(\hat{\bm{\beta}})(-1)^{m'}\threeJ{4}{2}{2}{m}{-m'}{n}\Fnu^{(22)}(T_0)a^{T}_{4 m}\\
		\end{cases}\\
		\text{Aberration} 
		&\begin{cases}
		+\beta \sum\nolimits_{m,M}^{1,1} D^{1}_{Mm}(\betahat)\CEu^{(1,+1)}_{1 M}D^{2}_{m'M}(\betahat^{-1})
		\Fnuc(T_0)a^{T}_{1 m}\\
		+\beta \sum\nolimits_{m,M}^{3,2} D^{3}_{Mm}(\betahat)\CEu^{(1,-1)}_{3 M}D^{2}_{m'M}(\betahat^{-1})
		\Fnuc(T_0)a^{T}_{3 m}\\
		+\frac{1}{2}\beta^2 \sum\nolimits_{m,M}^{2,2} D^{2}_{Mm}(\betahat)\CEu^{(2,0)}_{2 M}D^{2}_{m'M}(\betahat^{-1})
		\Fnuc(T_0)a^{T}_{2 m}\\
		+\frac{1}{2}\beta^2 \sum\nolimits_{m,M}^{4,2} D^{4}_{Mm}(\betahat)\CEu^{(2,-2)}_{2 M}D^{2}_{m'M}(\betahat^{-1})
		\Fnuc(T_0)a^{T}_{4 m}
		\end{cases}\\
		\text{Doppler+Aberration}
		&\begin{cases}
		+\beta^2 \sum\nolimits_{m,n,M',M}^{2,1,1,1} \frac{\sqrt{8\pi}}{\sqrt{3}}Y^*_{1n}(\betahat)D^{2}_{Mm}(\betahat)\CEu^{(1,-1)}_{2 M}D^{1}_{M'M}(\betahat^{-1})
		(-1)^{m'}\threeJ{1}{2}{1}{M'}{-m'}{n} \Fnuc^{(11)}a^{T}_{2 m}\\
		-\beta^2 \sum\nolimits_{m,n,M',M}^{2,1,3,2} 2\sqrt{\pi}Y^*_{1n}(\betahat)D^{2}_{Mm}(\betahat)\CEu^{(1,+1)}_{2 M}D^{3}_{M'M}(\betahat^{-1})
		(-1)^{m'}\threeJ{3}{2}{1}{M'}{-m'}{n} \Fnuc^{(11)}a^{T}_{2 m}\\
		-\beta^2 \sum\nolimits_{m,n,M',M}^{4,1,3,3} 2\sqrt{\pi}Y^*_{1n}(\betahat)D^{4}_{Mm}(\betahat)\CEu^{(1,-1)}_{4 M}D^{3}_{M'M}(\betahat^{-1})
		(-1)^{m'}\threeJ{3}{2}{1}{M'}{-m'}{n} \Fnuc^{(11)}a^{T}_{4 m}\\
		+O(\beta^3).\\
		\end{cases}
		\end{split}
		\end{equation}

\section{\label{sec:appC}Abbreviations}
In order to easily identify each intensity and polarization effect induced in the direction of a moving cluster, we used different names for individual terms. Table \ref{tab:1} contains a list of all these effect, including the equation they are defined in and a short description. 

\newcolumntype{M}{>{\centering\arraybackslash}m{\dimexpr.7\linewidth-1\tabcolsep}}
\begin{table*}[t]
	\centering
	\caption{}
	\renewcommand{\arraystretch}{1.5}
	\begin{tabular}{>{\centering\arraybackslash}m{\dimexpr.1\linewidth-2\tabcolsep} >{\centering\arraybackslash}m{\dimexpr.2\linewidth-2\tabcolsep} M}
		\hline
		Term & Definition & Description\\
		\hline
		\hline
		TinPol & Eq. \eqref{TinPol}, \eqref{TinPol2} \& \eqref{newTinPol}& Polarization distortion induced by the observed quadrupole ($a^{I_c}_{2 m}$) in the cluster's frame.\\
		\hline
		kSZPol & Eq. \eqref{kSZPol} \& \eqref{kSZPol2} & Polarization distortion induced by the leakage of the CMB rest frame monopole ($a^{I}_{00}$) into the observed quadrupole by a moving cluster. \\
		\hline
		T$_1$inPol & Eq. \eqref{T1inPol} & Polarization distortion induced by the leakage of the CMB rest frame dipole ($a^{I}_{1m}$) into the observed quadrupole by a moving cluster. \\
		\hline
		T$_2$inPol & Eq. \eqref{T2inPol} & Polarization distortion induced by the CMB rest frame quadrupole ($a^{I}_{2 m}$) in the cluster's frame. This effect is identical to TinPol for a non-moving cluster.\\
		\hline
		T$_3$inPol & Eq. \eqref{T3inPol} & Polarization distortion induced by the leakage of the CMB rest frame octupole ($a^{I}_{3m}$) into the observed quadrupole by a moving cluster. \\
		\hline
		T$_{1,3}$inPol & Eq. \eqref{T1inPol} \& \eqref{T3inPol} & Sum of T$_1$inPol and T$_3$inPol.\\
		\hline\hline
		bSZ$^{(\ell)}$ & Eq. \eqref{bSZ} \& \eqref{bSZ2} & Intensity distortion induced by the observed $\ell^{\text{th}}$ ($\ell\neq2$) multipole in the cluster's frame. This effect is entirely induced by scattering of the CMB photons out of the observer's line of sight. \\
		\hline
		kSZIn$^{(1)}$ & Eq. \eqref{kSZIn1} & Intensity distortion induced by the leakage of the CMB rest frame monopole ($a^{I}_{00}$) into the observed bSZ$^{(1)}$ in the cluster's frame.\\
		\hline
		T$_1$inbSZ$^{(1)}$ & Eq. \eqref{T1inbSZ} & Intensity distortion induced by the CMB rest frame dipole ($a^{I}_{1 m}$) in the cluster's frame. This effect is identical to bSZ$^{(1)}$ for a non-moving cluster. \\
		\hline
		T$_2$inbSZ$^{(1)}$ & Eq. \eqref{T2inbSZ} & Intensity distortion induced by the leakage of the CMB rest frame monopole ($a^{I}_{2m}$) into the observed bSZ$^{(1)}$ in the cluster's frame.\\
		\hline\hline
		TinIn & Eq. \eqref{TinIn}, \eqref{TinIn2} \& \eqref{newTinIn} & Intensity distortion induced by the observed quadrupole ($a^{I_c}_{2 m}$) in the cluster's frame. This term is essentially bSZ$^{(2)}$ which has been named differently---in accordance with TinPol---to distinguish it from the rest of the bSZ terms.  \\
		\hline
		kSZIn$^{(2)}$ & Eq. \eqref{kSZIn2} & Intensity distortion induced by the leakage of the CMB rest frame monopole ($a^{I}_{00}$) into the observed quadrupole in the cluster's frame.\\
		\hline
		T$_1$inIn & Eq. \eqref{T1inIn} & Intensity distortion induced by the leakage of the CMB rest frame dipole ($a^{I}_{1m}$) into the observed quadrupole by a moving cluster.\\
		\hline
		T$_2$inIn & Eq. \eqref{T2inIn} & Intensity distortion induced by the CMB rest frame quadrupole ($a^{I}_{2 m}$) in the cluster's frame. This effect is identical to TinIn for a non-moving cluster.\\
		\hline
		T$_3$inIn & Eq. \eqref{T3inIn} & Intensity distortion induced by the leakage of the CMB rest frame dipole ($a^{I}_{1m}$) into the observed quadrupole by a moving cluster.\\
		\hline
		T$_{1,3}$inIn & Eq. \eqref{T1inIn} \& \eqref{T3inIn} & Sum of T$_1$inIn and T$_3$inIn.\\
		\hline
		kSZIn & Eq. \eqref{kSZIntot} & Sum of kSZIn$^{(1)}$ and kSZIn$^{(2)}$.\\
		\hline		
	\end{tabular}
	\label{tab:1}
\end{table*}
		
\end{widetext}
\end{document}